\documentclass[a4paper,usenatbib]{mnras}

\makeatletter

\newcommand{\Rmnum}[1]{\expandafter\@slowromancap\romannumeral #1@}

\newcommand{\gsim}{\lower0.6ex\vbox{\hbox{$\buildrel{\textstyle >}\over{\sim}\ $}}}

\makeatother

\usepackage{newtxtext,newtxmath}
\usepackage{graphicx}	
\usepackage{amsmath}	
\usepackage{subcaption}
\usepackage{booktabs}
\usepackage[dvipsnames,svgnames]{xcolor}
\usepackage{color}
\def\astrid{\texttt{Astrid}}
\def\bluetides{\texttt{BlueTides~}}

\def\hmsun{{h^{-1} M_{\odot}}}
\def\msun{\, M_{\odot}}

\title[ASTRID MBH Mergers]
{Massive Black Hole Mergers with Orbital Information: Predictions from the ASTRID Simulation}

\author[N.~Chen et al.]{
Nianyi Chen,$^{1}$\thanks{E-mail: nianyic@andrew.cmu.edu}
Yueying Ni,$^{1,2}$
A.~Miguel Holgado,$^{1}$
Tiziana Di Matteo,$^{1,2}$
Michael Tremmel,$^{3}$
Colin DeGraf,$^{4}$
\newauthor
Simeon Bird,$^{5}$
Rupert Croft,$^{1,2}$
Yu Feng$^{6}$
\\
$^{1}$ McWilliams Center for Cosmology, Department of Physics, Carnegie Mellon University, Pittsburgh, PA 15213 \\
$^{2}$ NSF AI Planning Institute for Physics of the Future, 
Carnegie Mellon  University, Pittsburgh, PA 15213, USA \\
$^{3}$ Astronomy Department, Yale University, P.O. Box 208120, New Haven, CT 06520, USA \\
$^{4}$ Department of Physics, Truman State University, Kirksville, MO 63501, USA \\
$^{5}$ Department of Physics \& Astronomy, University of California, Riverside, 900 University Ave., Riverside, CA 92521, USA\\
$^{6}$ Berkeley Center for Cosmological Physics and Department of Physics, University of California, Berkeley, CA 94720, USA
}

\date{Accepted XXX. Received YYY; in original form ZZZ}

\pubyear{2021}
\begin{document}
\maketitle

\begin{abstract}
We examine massive black hole (MBH) mergers and their associated gravitational wave signals from the large-volume cosmological simulation \astrid~.
\astrid~includes galaxy formation and black hole models recently updated with a MBH seed population between $3\times 10^4\hmsun$ and $3\times 10^5\hmsun$ and a sub-grid dynamical friction (DF) model to follow the MBH dynamics down to $1.5\;\text{ckpc}/h$.
We calculate initial eccentricities of MBH orbits directly from the simulation at kpc-scales, and find orbital eccentricities above $0.7$ for most MBH pairs before the numerical merger. 
After approximating unresolved evolution on scales below ${\sim 200\,\text{pc}}$, we find that the in-simulation DF on large scales accounts for more than half of the total orbital decay time ($\sim 500\,\text{Myrs}$) due to DF.
The binary hardening time is an order of magnitude longer than the DF time, especially for the seed-mass binaries ($M_\text{BH}<2M_\text{seed}$).
As a result, only $\lesssim20\%$ of seed MBH pairs merge at $z>3$ after considering both unresolved DF evolution and binary hardening. 
These $z>3$ seed-mass mergers are hosted in a biased population of galaxies with the highest stellar masses of $>10^9\,M_\odot$.  With the higher initial eccentricity prediction from \astrid~, we estimate an expected merger rate of $0.3-0.7$ per year from the $z>3$ MBH population.
This is a factor of $\sim 7$ higher than the prediction using the circular orbit assumption.
The LISA events are expected at a similar rate, and comprise $\gtrsim 60\%$ seed-seed mergers, $\sim 30\%$ involving only one seed-mass MBH, and $\sim 10\%$ mergers of non-seed MBHs.




\end{abstract}


\section{Introduction}
\label{sec:intro}
Massive Black Holes (MBHs) are known to exist at the center of galaxies \citep[e.g.][]{Soltan1982,Kormendy1995,Magorrian1998,Kormendy2013}. 
As these galaxies merge \citep[e.g.][]{Lacey1993,Lotz2011,Rodriguez-Gomez2015}, the MBHs that they host will also merge, resulting in the mass growth of the MBH population \citep[e.g.][]{Begelman1980}.
MBH mergers following their host galaxy mergers are an important aspect of the growth for MBHs in dense environments \citep[e.g.][]{Kulier2015}. 
Even more importanly, as a by-product of MBH mergers, gravitational waves are emitted, and their detection opens up a new channel for probing the formation and evolution of early MBHs in the universe \citep[e.g.][]{Sesana2007a,Barausse2012}.

The gravitational wave (GW) detection by LIGO \citep[][]{LIGO2016PhRvL.116f1102A} proves the experimental feasibility of using gravitational waves for studying black hole (BH) binaries. 
While LIGO cannot detect GWs from binaries more massive than $\sim 100 M_\odot$ \citep[][]{Mangiagli2019}, long-baseline experiments are being planned for detections of more massive BH binaries. 
Specifically, the upcoming Laser Interferometer Space Antenna (LISA) \citep{LISA2017arXiv170200786A} mission will be sensitive to low-frequency ($10^{-4}-10^{-1}$Hz) gravitational waves from the coalescence of MBHs with masses $10^4-10^7 M_\odot$ up to $z\sim 20$. 
 At lower frequencies, Pulsar Timing Arrays (PTAs) are already collecting data and the
Square Kilometer Array (SKA) in the next decade will be a major leap forward in
sensitivity. 
While MBH binaries are the primary sources for PTAs and LISA, these two experiments probe different stages of MBH evolution. 
PTAs are most sensitive to the early inspiral
(orbital periods of years or longer) of nearby ($z <1$) (massive) sources  \citep{Mingarelli2017}. 
In contrast, LISA is sensitive to the inspiral, merger, and ringdown of MBH binaries  at a wide range of redshifts \citep{Amaro-Seoane2012}. 

GWs from MBH mergers will provide a unique way of probing the high-redshift universe and understanding the early formation of the MBH seeds, especially when combined with observations of the electromagnetic (EM) counterparts \citep[][]{Natarajan2017,DeGraf2020}.
For instance, a MBH merger multi-messenger detections should allow us to distinguish between different MBH seeding mechanisms at high-redshift \citep[][]{Ricarte2018}, to obtain information on the dynamical evolution of massive black holes \citep[][]{Bonetti2019}, and to gain information about the gas properties within the accretion disc \citep[][]{Derdzinski2019}.

To properly access the potential of the upcoming GW signals as well as the EM observations of  MBH binaries, we need to gain a thorough understanding of the physics of these events with theoretical tools and be able to make statistical predictions for the binary population. 
Early studies have provided merger rate predictions for MBH binaries using analytic models \citep[e.g.][]{Haehnelt1994,Jaffe2003,Wyithe2003}. 
Some more recent predictions made use of semi-analytic models \citep[e.g.][]{Sesana2004,Tanaka2009,Barausse2012,Ricarte2018} to enhance the model complexity and physical realism.
Recent developments in large-volume cosmological simulations \citep[e.g.][]{Hirschmann2014,Vogelsberger2014,Schaye2015,Feng2016,Volonteri2016,Pillepich2018,Dave2019} have enabled the study of MBH mergers within the context of cosmological galaxy formation \citep[e.g.][]{Salcido2016,Kelley2017a,Katz2020,Volonteri2020}. 
These simulations directly associate MBH binaries with their host galaxies, and they are carried out in large enough cosmological volumes to provide the statistical power to make merger rate predictions across cosmic time which are crucial for the upcoming observations. 

In order to accurately predict when MBH mergers occur in these simulations, one must account for the orbital decay and binary hardening timescales in a wide dynamical range. 
During galaxy mergers, the central MBHs start at large separation in the remnant galaxy (as much as a few tens of kpc). 
These MBHs then gradually lose their orbital energy and sink to the center of the remnant galaxy due to the dynamical friction exerted by the gas, stars and dark matter around them \citep[e.g.][]{Chandrasekhar1943,Ostriker1999}. 
When their separation is $\lesssim 1$ parsec, a MBH binary forms and other energy-loss channels begin to dominate, such as scattering with stars \citep[e.g.][]{Quinlan1996,Sesana2007b,Vasiliev2015}, gas drag from the circumbinary disk \citep[e.g.][]{Haiman2009}, or, if relevant, three-body scattering with a third black hole \citep[e.g.][]{Bonetti2018}.

Among these processes, only the dynamical friction decay affects the dynamics at orbital separation above the resolution of large-volume cosmological simulations. 
However, so far there is limited attempt to directly model dynamical friction (at small scales, close to the resolution) in the large-volume cosmological simulations mentioned above. 
In most cosmological simulations, once MBHs are within a given halo, they are simply repositioned to the minimum potential position of the host galaxy at each time-step. 
For these simulations, (although sometimes the effects of subgrid dynamical friction are treated in post-processing), many spurious mergers occur during fly-by encounters. 
Among simulations that do include subgrid modeling of DF on-the-fly, \citet{Dubois2014} only includes the friction from gas but not stars, while \citet{Tremmel2017} and \citet{Hirschmann2014} model the dynamical friction from stars and dark matter particles. 

Here we study MBH mergers using the large volume cosmological simulation \astrid~which
uses a novel power-law seeding with a range of MBH seed masses and so includes relatively low mass MBHs. 
More importantly, it directly incorporates additional dynamical friction modeling,  following the recent model by 
\citep[][]{Chen2021} for the MBH dynamics down to the resolution limit \citep[see also similar implementations by][]{Hirschmann2014,Tremmel2015}.
With more physical modeling of the MBH dynamics, we can follow the in-simulation mergers for a more extended period of time over hundreds of Myrs, and almost completely prevent mergers during fly-by encounters. 
Moreover, for the first time we can aim to measure the orbital evolution and eccentricities of MBH pairs on sub-kpc scales. 
Such information should be important both for estimating the binary hardening timescales, and for predicting the GW signals from the MBH mergers. 

This paper is organized as follows: in Section \ref{sec:sim} we introduce the \astrid~simulation, in particular the MBH modeling, and describe how we obtain the merger catalog from the simulation; in Section \ref{sec:ecc}, we describe our methods for measuring the MBH orbital eccentricity from the simulation, and present results of our measurements. 
Section \ref{sec:delay_models} focuses on the modeling of post-processing delay times including the dynamical friction time and binary hardening time after the numerical merger. 
Then in Section \ref{sec:rate}, we present our prediction for MBH merger rate at $z>3$, and investigate the properties of high-redshift MBH merger systems.
Finally, in Section \ref{sec:snr} we show the GW strain and signal-to-noise ratios for the binary population that merges at $z>3$.

\section{Simulation}
\label{sec:sim}
\subsection{The \astrid~Simulation}
\label{subsec:asterix}

The \astrid~simulation is a large-scale cosmological hydrodynamic simulation in a $250\, {\rm Mpc}/h$ box with $2\times 5500^3$ particles. 
\astrid~contains a statistical sample of halos which can be compared to future survey data from JWST, while resolving galactic halos down to $10^9 \msun$ (corresponding to 200 dark matter particles). 
\astrid~has been run from $z=99$ to $z=3$. 
It contains models for inhomogeneous hydrogen and helium reionization, baryon relative velocities and massive neutrinos, as well as 'full-physics' star formation model, BH accretion and associated supernova and AGN feedback respectively. 
The BH model includes mergers driven by dynamic friction rather than repositioning. 
Our treatment of MBHs largely follows the \bluetides~simulation in terms of the BH accretion and feedback, which is based on the earlier work by \cite{DiMatteo2005,Springel2005b}.
Compared with \bluetides, we slightly changed the seeding scheme of MBHs by drawing the seed mass from a power-law distribution instead of using a universal seed mass.
Furthermore, we use a dynamical friction model \citep[tested and validated in][]{Chen2021} to evolve the binary black holes and include the sinking and merger of MBHs in the simulation in a more physical way.
Here we briefly summarize the black hole seeding and dynamics treatment in \astrid, and refer to \cite{Bird2021} and \cite{Ni2021} for detailed presentations of physical models for star formation and black holes.

\subsubsection{MBH Seeding}

To seed MBHs in the simulation, we periodically run a FOF group finder on the fly with a linking length of 0.2 times the mean particle separation, to identify halos with a total mass and stellar mass satisfying the seeding criteria \{ $M_{\rm halo,FOF} > M_{\rm halo,thr}$; $M_{\rm *,FOF} > M_{\rm *,thr}$\}.
We apply a mass threshold value of $M_{\rm halo,thr} = 5 \times 10^9 \hmsun$ and $M_{\rm *,thr} = 2 \times 10^6 \hmsun$.

Considering the complex astrophysical process involved in BH seed formation in realistic cases, halos with the same mass can form different mass MBH seeds.
Therefore, in \astrid, instead of applying a uniform seed mass for all MBHs, we probe a mass range of the MBH seed mass $M_{\rm seed}$ drawn probabilistically from a power-law distribution:
\begin{equation}
\label{equation:power-law}
    P(M_{\rm seed}) = 
    \begin{cases}
    0 & M_{\rm seed} < M_{\rm seed,min} \\
    \mathcal{N} (M_{\rm seed})^{-n} & M_{\rm seed,min} \leq M_{\rm seed} \leq M_{\rm seed,max} \\
    0 & M_{\rm seed} > M_{\rm seed,max}
   \end{cases}
\end{equation}
where $\mathcal{N}$ is the normalization factor.
The minimum seed mass is $M_{\rm seed,min} = 3 \times 10^4 \hmsun$ and the maximum seed mass is $M_{\rm seed,max} = 3 \times 10^5 \hmsun$, with a power-law index $n = -1$. 
For each halo that satisfies the seeding criteria but does not already contain at least one BH particle, we convert the densest gas particle into a BH particle. 
The new-born BH particle inherits the position and velocity of its parent gas particle. 

\subsubsection{MBH Dynamics}
Instead of constantly repositioning the black hole towards the potential minimum, as in earlier simulations, in \citet{Chen2021} we implemented and tested a model for sub-grid dynamical friction \citep[similar to][]{Tremmel2015, Tremmel2017}. 
Dynamical friction is an artificial force for modelling unresolved small-scale interactions between the MBH and nearby stars and dark matter. 
These interactions transfer momentum from the MBH to individual stars in the surrounding star clusters,  gradually reducing the momentum of the MBH particle relative to the surrounding collisionless objects in the bulge \citep[e.g.][]{Governato1994,Kazantzidis2005}. 
The additional dynamical friction also stabilizes the MBH motion at the center of the galaxy.

We estimate dynamical friction on MBHs using Eq. 8.3 of \cite{Binney2008}: 
\begin{equation}
\label{eq:df_full}
    \mathbf{F}_{\rm DF} = -16\pi^2 G^2 M_{\rm BH}^2 m_{a} \;\text{log}(\Lambda) \frac{\mathbf{v}_{\rm BH}}{v_{\rm BH}^3} \int_0^{v_{\rm BH}} dv_a v_a^2 f(v_a),
\end{equation}
where $M_{\rm BH}$ is the BH mass, $\textbf{v}_{\rm BH}$ is the BH velocity relative to its surrounding medium, $m_a$ and $v_a$ are the masses and velocities of the particles surrounding the BH, and $\text{log}(\Lambda)=\text{log}(b_{\rm max}/b_{\rm min})$ is the Coulomb logarithm that accounts for the effective range of the friction between the specified $b_{\rm min}$ and $b_{\rm max}$. 
$f(v_a)$ in Eq.~\ref{eq:df_full} is the velocity distribution of the surrounding collisionless particles including both stars and dark matter. 
Here we have assumed an isotropic velocity distribution of the particles surrounding the BH so that we are left with a 1D integration.

In \astrid, the BH seed mass extends down to $3\times 10^4 M_\odot/h$, which is one order of magnitude smaller than the stellar particle mass. 
In this regime, the dynamical friction of BH is likely unrealistic due to its small mass compared to the masses of other particles, and so the dynamics of the seed BH would be unstable due to dynamical heating (when $M_{\rm BH}$ is below the mass resolution). 
Therefore, we boost the dynamical friction in this regime with
$M_{\rm dyn}=2\times M_{\rm DM}$ when
 $M_{\rm BH} < M_{\rm dyn}<1$. 
 This temporarily boosts the BH dynamical mass for BHs near the seed mass and helps stabilize their motion during the early post-seeding evolution.

We approximate the distribution function $f(v_a)$ by the Maxwellian distribution \citep[as, e.g.][]{Binney2008}, and account for the neighbouring collisionless particles up to the range of the SPH kernel of the BH particle \citep[see,][for more details]{Chen2021}. 
Eq.~\ref{eq:df_full} becomes
\begin{equation}
    \label{eq:H14}
    \mathbf{F}_{\rm DF} = -4\pi \rho_{\rm sph} \left(\frac{GM_{\rm dyn}}{v_{\rm BH}}\right)^2  \;\text{log}(\Lambda) \mathcal{F}\left(\frac{v_{\rm BH}}{\sigma_v}\right) \frac{\bf{v}_{\rm BH}}{v_{\rm BH}}.
\end{equation}
Here $\rho_{\rm sph}$ is the density of dark matter and star particles within the SPH kernel,
$\mathcal{F}$ is the integral in Equation \ref{eq:df_full} assuming a Maxwellian distribution of stellar velocities. $\sigma_v$ is the velocity dispersion of the dark matter and star particles within the SPH kernel.

The boost of the initial $M_{\rm dyn}$ may overestimate the dynamical friction for small BHs and the resultant sinking timescale will be shortened by a factor of $\sim M_{\rm BH}/M_{\rm dyn}$ compared to the no-boost case.
On the other hand, it is also possible that the BH sinking time scale estimated in our simulation in the no-boost case could overestimate the true sinking time, as the high-density stellar bulges are not fully resolved \citep[e.g.][]{Antonini2012,Dosopoulou2017,Biernacki2017}.
Therefore, boosting the initial $M_{\rm dyn}$ seems a reasonable compromise to model the dynamics of small mass BHs while also alleviating the noisy perturbation of dynamic heating brought by the limit of resolution. 
Note that even if our dynamic friction implementation overestimates the force, it still provides a substantially more conservative estimation of BH sinking than the common model where BHs are repositioned to the potential minimum.

\subsection{Merger Catalog}
\label{subsec:catalog}

There are a total of 445635 BH mergers in the simulation for $z> 3$. 
For each merger event we extract the relevant environmental variables (the density profiles of gas, dark matter and stars, and the stellar velocity dispersion) from the nearest snapshot before and after the merger. 
The snapshots used are separated by $\sim 20$ Myrs. 
In a small fraction of cases, the mergers take place within $20\,{\rm Myrs}$ after one of the MBHs are born, and so we cannot find the corresponding MBH in the previous snapshot. 
We remove these mergers from the catalog, after which 440999 mergers remain. 

From the snapshots immediately before and after the merger, we identify the host halos and subhalos containing the binaries using FOF and SubFind, respectively. 
Out of the mergers that remain in the catalog, we further remove those not associated with any halo/subhalo, and those whose host galaxy has less than $200$ star particles.
The hosts for these binaries are not well resolved in our simulation, so we cannot reliably compute the binary hardening time in post processing. 
This leaves us with a final catalog of 430938 black hole merger candidates.
For each host halo identified, we define the halo center as the position of the particle with the minimum potential, and calculate galaxy properties such as the density profiles and half-mass radius with respect to this point.

In Figure \ref{fig:image}, we show the last few orbits of a few selected BH pairs in our merger catalog plotted on their host galaxies' stellar distributions. 
The distance from left to right of each image is $8\,{\rm ckpc}/h$.
The brightness corresponds to the stellar density, and the colours show the stellar age with older stars being redder and younger stars being whiter. 
The red curves are the BH pairs' positions relative to their center of mass.

\section{Orbital Eccentricity}
\label{sec:ecc}
\begin{figure*}
    \includegraphics[width=0.99\textwidth]{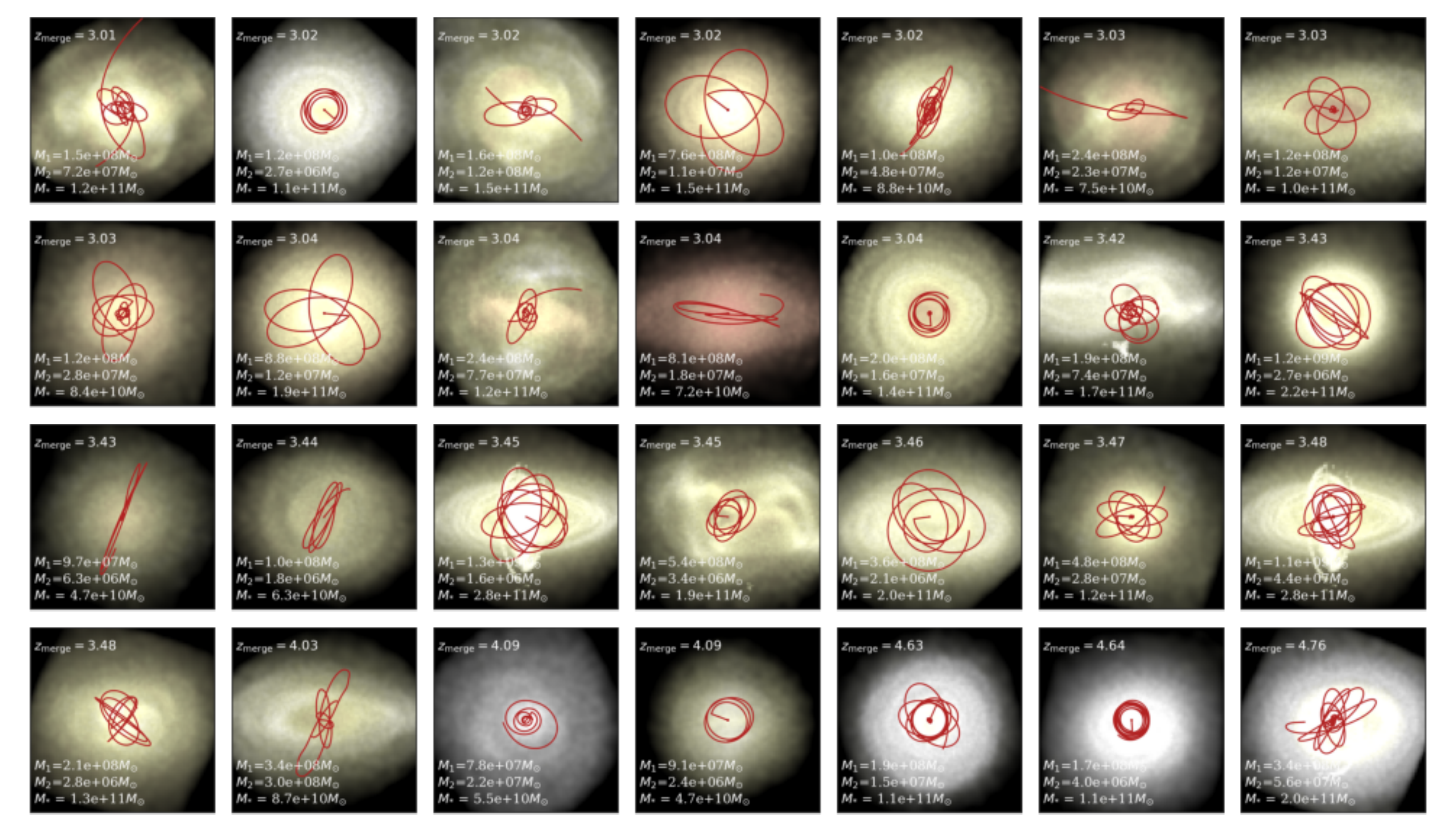}
    \caption{The last few orbits (starting from $\sim 80\,{\rm Myrs}$ before the merger) of selected binaries in the \astrid~simulation plotted on their host galaxies. The distance from left to right of each image is $8\,{\rm ckpc}/h$. The brightness corresponds to the stellar density, and the colours show the stellar age with older stars being redder. The red curves are the BH pairs' position relative to their center of mass. In most cases we see a Rosetta orbit, as the local potential is a spherical potential dominated by stars and dark matter. We find that some orbits circularize over time (e.g. third row, fifth column), although the majority of the orbits still remain eccentric when merging (see e.g. Figure \ref{fig:ecc}).}
    \label{fig:image}
\end{figure*}

\begin{figure*}
    \includegraphics[width=0.99\textwidth]{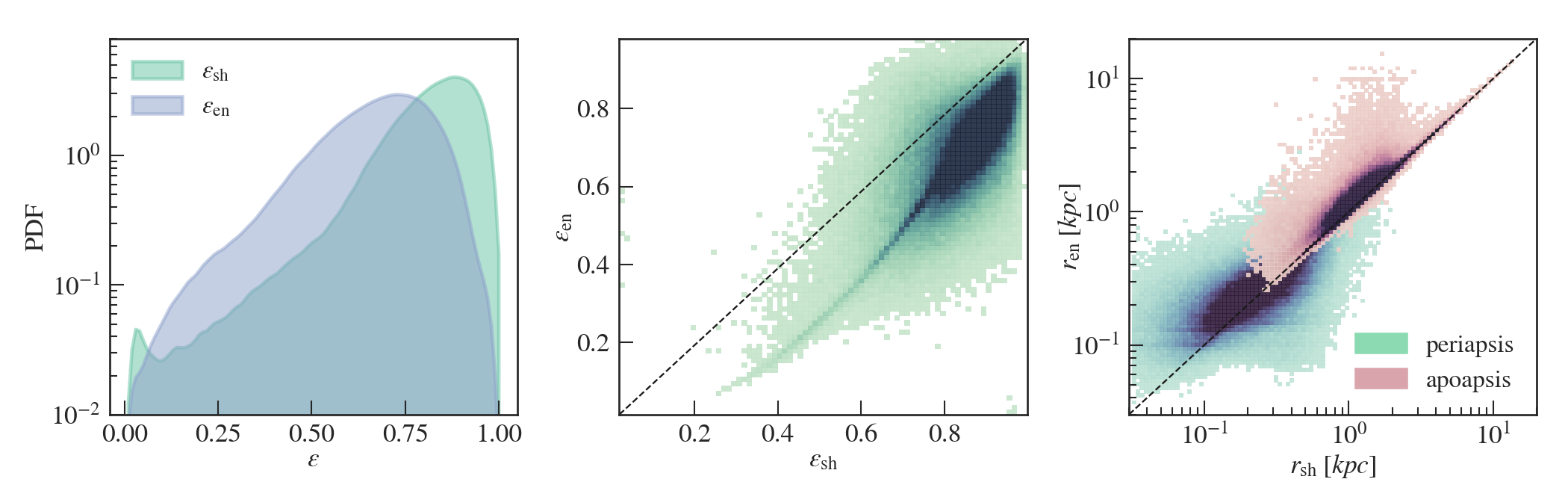}
    \caption{Comparison between eccentricity measurements from the shape method and the energy method. \textbf{Left:} the distribution of the (generalized) orbital eccentricity from the two measurements. In both cases, the distribution is dominated by highly-eccentric binaries, as we can also see from the images in Figure \ref{fig:image}. The shape method has a more skewed distribution compare to the energy method. \textbf{Middle:} A scattered plot of the eccentricity from the two measurements. We can see that the two measurements yield similar results by comparing the distribution to the diagonal line. In most cases, the energy measurement is $\sim 10\%$ lower than the shape measurement. \textbf{Right:} In addition to the eccentricity, we show the apoapses and periapases of the two measurements. The orange dots are the apoapses and the green dots are the periapases. The scatter relation also follows the diagonal line quite closely. When the two black holes merge in the simulation, the apoapsis is usually a few kpc and the periapsis is usually less than 1kpc.}
    \label{fig:ecc}
\end{figure*}

As was described in Section \ref{subsec:asterix}, our simulation has a build-in sub-grid dynamical friction model, which allows us to follow the black holes' orbits before their numerical mergers down to the resolution limit. 
Figure \ref{fig:image} shows several examples of the last few orbits of BH pairs just before they merge in the simulation. 
The black hole orbits are plotted in the center-of-mass frame of the BH pairs, with a face-on projection on the 2D plane perpendicular to the mean angular momentum of the last orbit. 
Since we record the BH information at each time step when the BH is active, the orbits are much better resolved in time compared with the galaxies. 
Most orbits start off at a semi-major axis of $>1$ kpc, and gradually go through orbital decay until merger. 

From the images, we see that the majority of the orbits are very non-circular during the initial encounter of the BHs. 
While some of them circularize with time, most orbits still retain a high eccentricity at the time of merger in the simulation. 
This motivates us to characterize the orbital eccentricity before merging, as it is an important piece of information not only for estimating the binary hardening time with analytical models, but also for calculating the GW signals from the merger events. 
In this section, we will describe two ways of characterizing the orbital eccentricities of the BH pairs in our simulation.\footnote{We also tried applying the method of osculating elements \citep[e.g.,][and references therein]{Efroimsky2004} to the orbital trajectories; however, we found that the stellar environment dominated the binary's evolution, such that it could not be adequately described as a post-Keplerian orbit.}

\subsection{Shape Measurement}
\label{sec:shape}
Given the images in Figure \ref{fig:image}, a natural way of measuring the orbital eccentricity is to use the shape of the orbits just before the numerical merger, and this is the first approach we take.

On $\sim {\rm kpc}$ scales, since most orbits are not Keplerian except those of the most massive BHs and the orbits are constantly shrinking, the BH orbits do not fit an ellipse. 
Instead, they exhibit the feature of a Rosetta orbit (the feature is most prominent in e.g. second row, second column of Figure \ref{fig:image}, although standard Rosetta orbits do not shrink over time). 
For orbits resulted from the spherically symmetric potential, we can characterize the eccentricity by the size of the inner radius and the size of the outer radius. More specifically, for each orbit, we define $\Delta r_2$ to be position of the secondary BH with respect to the center of mass, and we take the local minimum of $\Delta r_2$ as the (generalized) periapsis of the orbit, and the local maximum of $\Delta r_2$ as the apoapsis. 
Then, we represent the orbital eccentricity of the binary by the generalized eccentricity, defined for a spherical potential as:
 \citep{Binney2008}:
\begin{equation}
\label{eq:gen_e}
    \epsilon = \frac{r_{\rm apo} - r_{\rm peri}}{r_{\rm apo} + r_{\rm peri}},
\end{equation}
where $r_{\rm apo}$ and  $r_{\rm per}$ are the apoapsis and the periapsis of the orbit, respectively. 
To distinguish between the measurement of the two methods, we will use the subscript "sh" to refer to the measurements from this shape-based method.
We average the eccentricity measurements over the last three orbits. 
We note, however, that the distribution in eccentricity does not change significantly when we take the average of the last one, two or three orbits.

\subsection{Solving the Orbital Equation}
\label{sec:energy}
In addition to the shape-based measurement in \S3.1, we also calculate the generalized orbital eccentricity by simply solving the orbital equation. 
Using these two independent methods we will then be able to compare the robustness of the BH orbit eccentricity distribution measurement from the simulations.

When the BH merger occurs in the simulation, the separation between the black hole pair is $\sim 3$ ckpc/$h$. 
At this distance, the gravitational potential is dominated by the surrounding stars and dark matter particles instead of the BHs themselves. 
Under such circumstances, the orbit of the satellite BH is non-Keplerian, as we have shown in Section \ref{sec:shape}. 
In the case of a spherical potential, the (generalized) apoapsis and periapsis can be obtained by solving the generalized orbital equation \citep{Binney2008}:
\begin{equation}
\label{eq:orbit}
    \frac{1}{r^2} + \frac{2[\phi(r)-E]}{J^2} = 0,
\end{equation}
where $\phi(r)$ is the gravitational potential computed from the density profile of surrounding particles, $E$ is the total energy per unit mass and $J$ is the angular momentum per unit mass of the secondary black hole with respect to the host galaxy center. 
The larger root of the equation corresponds to $r_{\rm apo}$ and the smaller root is $r_{\rm peri}$. 

When solving Equation \ref{eq:orbit}, we take $E$ and $J$ to be the average energy and angular momentum over the last half-orbit (i.e. from the last local maximum to the last local minimum of the BH separation) of the BH. 
We did not take the average over a more extended period of time because the black hole pair is constantly losing energy. 
After getting the two apses, we again use Equation \ref{eq:gen_e} to calculate the orbital eccentricity. 
We refer to this method as the energy method, and use subscript "en" when showing results.

Figure \ref{fig:ecc} shows a comparison between the (generalized) eccentricity measurements from the shape method and the energy method. 
The left panel shows the distribution of eccentricities for all the mergers in the simulation. 
The measurements from both methods show that the BH binary  population dominated by highly eccentric orbits, with a peak at $\epsilon\sim 0.85$ for the shape-based method and $\sim 0.75$ for the energy-based method. 
Comparing the two distributions, we see that the shape measurement generally produces a distribution with higher eccentricities than the energy method.  
In the middle panel we show a scatter plot of the eccentricities from the two measurements. 
There is a positive correlation between the two eccentricities, with the majority of the measurements close to the diagonal line. 
This means that the two measurements are not only close in distribution, but also yield correlated results for each individual orbit. 
Similar to what shown by the 1D distributions, the shape method predicts higher values of eccentricity for most pairs than the energy method (typically $\sim$ 10\%  lower).

In addition to the eccentricity, in the right panel we further compare the apoapses and periapases from the two measurements. 
Overall, we can see that the apoapsis peaks around $1\sim 3$ kpc, while the periapsis peaks around $0.1\sim 0.7$ kpc. 
Again there is a good alignment between the two measurements, with the peaks distributed close to the diagonal line. 
In the majority of cases the shape measurement gives an larger apoapsis value.

To estimate the binary hardening time, we will use the measured binary eccentricities as an input to the model. 
By doing so, we do not consider any time-evolution of the binary eccentricity due to dynamical friction beyond the point of numerical merger \citep{Colpi1999,Hashimoto2003}.
In particular,  we will only show results using the values from the energy-based method ($\epsilon_{\rm en}$) in later sections, and we have tested that the effect of using the shape-based values is minor compared with the uncertainties from other sources (e.g. density in the central region of the galaxy).

\section{Post-processing Delays}
\label{sec:delay_models}

\subsection{Dynamical Friction}
\label{sec:df}

\begin{figure}
\centering
    \includegraphics[width=0.49\textwidth]{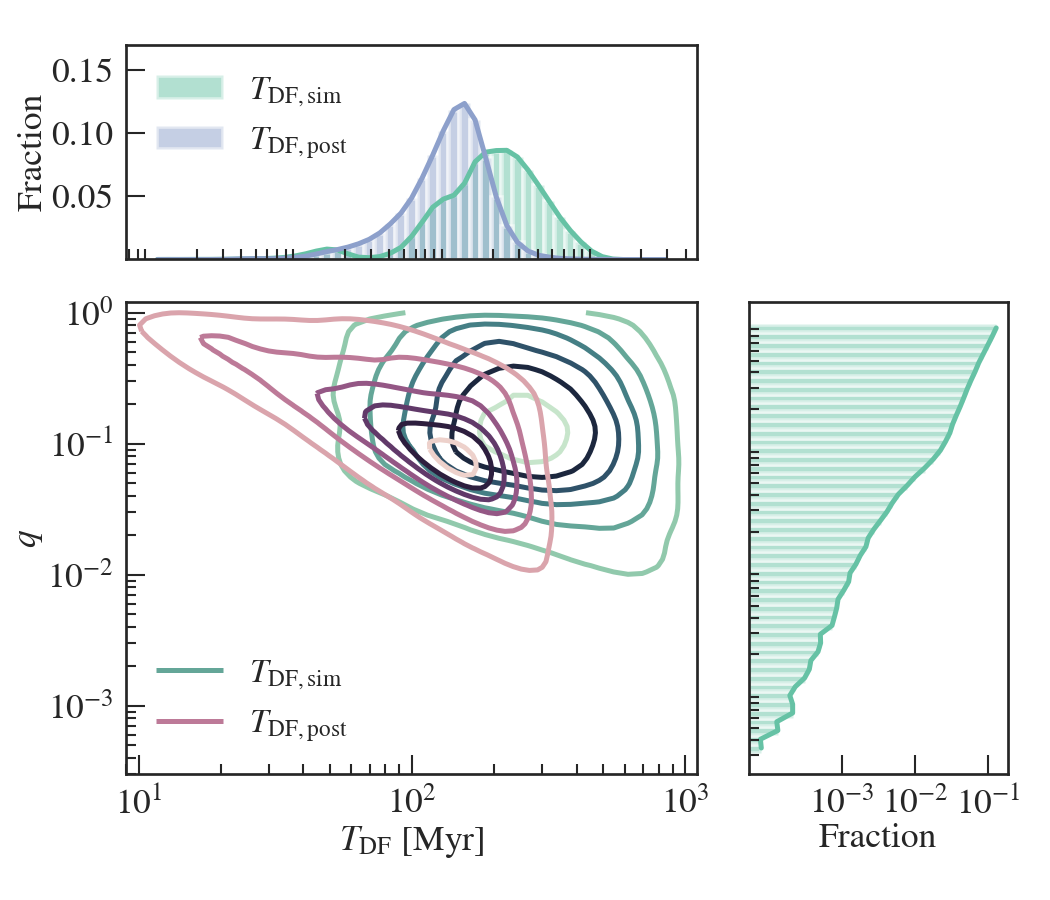}
    \caption{Comparison between the pre-merger dynamical friction time and the post-merger dynamical friction time. \textbf{Top:} Distributions of the pre- and post- merger DF times for all MBH pairs in \astrid. The two distributions are similar and both peaks around 200 Myrs, indicating that by adding dynamical friction to the simulation, we have resolved more than half of the total dynamical friction delay. \textbf{Bottom left:} Relation between the DF times and the mass ratio between the two MBHs ($q$). We observe the expected negative correlation between DF times and $q$. \textbf{Bottom right:} 1D distribution of the mass ratio $q$.}
    \label{fig:t_df}
\end{figure}

\begin{figure}
\centering
    \includegraphics[width=0.48\textwidth]{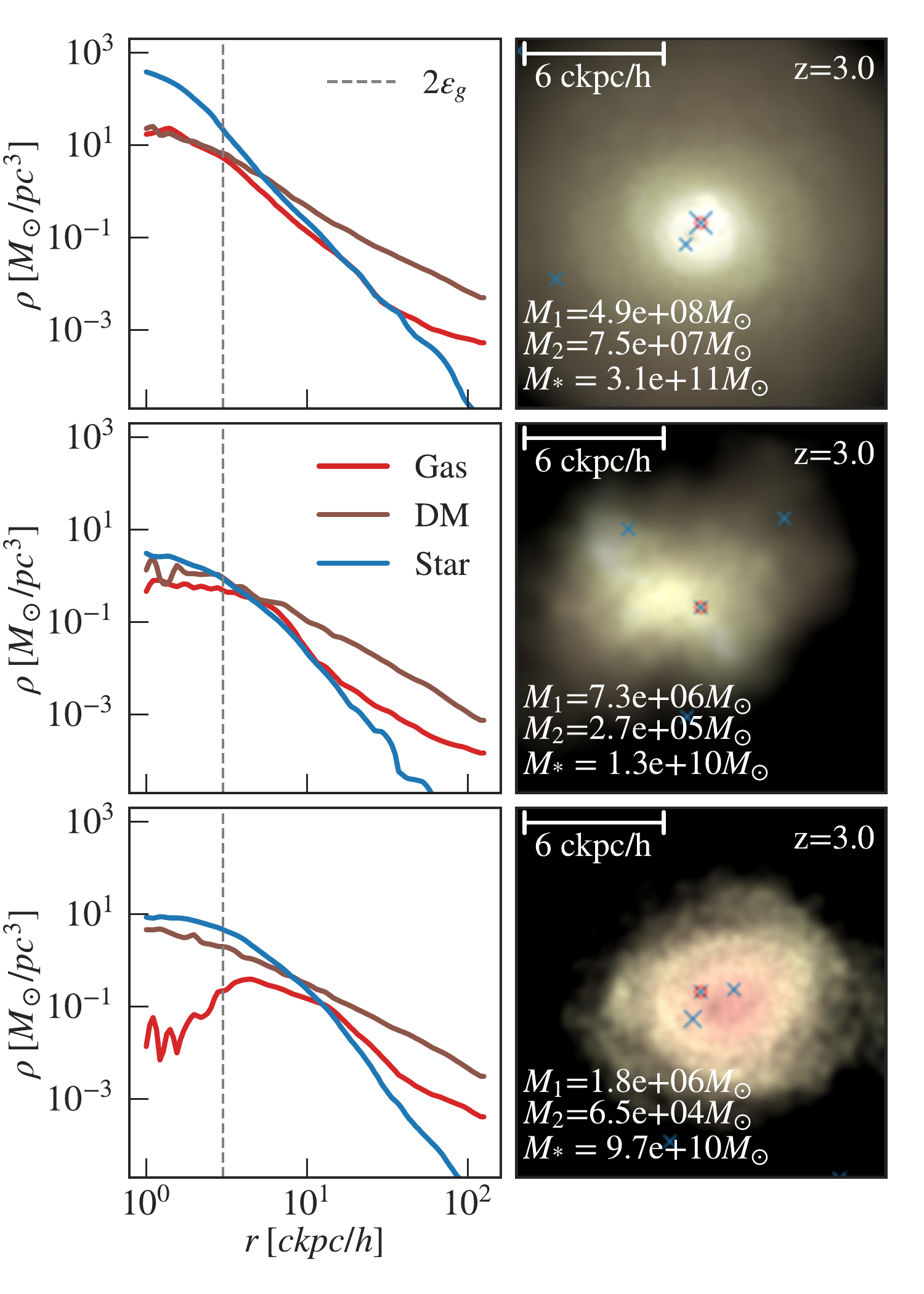}
    \caption{Density profiles (left) and images (right) of the host galaxies of three MBH mergers in the simulation. 
    The blue crosses mark all MBHs in the host galaxy, scaled by the BH mass. 
    The red circles mark the merging binary.
    \textbf{Top:} Host of a very massive binary with $M_{\rm tot}=5.6\times 10^8 M_\odot$ at $z=3$. 
    The stellar density is the dominant component on scales below $\sim 10$ ckpc/$h$.
    \textbf{Middle:} Host of a binary with $M_{\rm tot}=7.6\times 10^6 M_\odot$ at $z=3$. For this less massive binary, the density of the three components is comparable at $r<10$ ckpc$/h$, and the density profile flattens at a larger radius. 
    \textbf{Bottom:} Host of a binary with two seed-mass MBHs.
    The mass of the host galaxy is high relative to the binary mass.
    The binary is not the most massive MBHs in this galaxy, but the merger still occurs at a relatively central region.}
    \label{fig:prof}
\end{figure}

\begin{figure*}
    \includegraphics[width=0.99\textwidth]{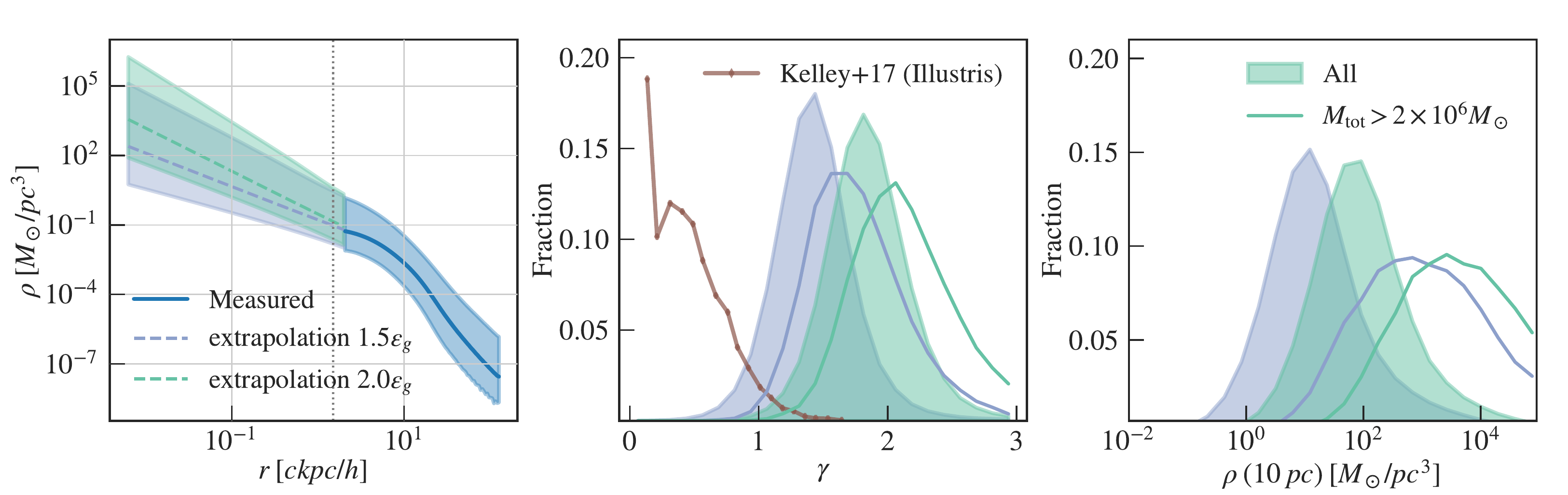}
    \caption{
    \textbf{Left:} The density profiles of \astrid~galaxies that host a recent numerical merger. The blue solid line shows the median density of all binary hosts measured from the simulation and the shaded region encloses 95\% of the population. The power law extrapolation is shown by dashed lines. 
    Here we show the results for extrapolation scales $r_{\rm ext} = 1.5\epsilon_g$ (purple) and $r_{\rm ext} = 2\epsilon_g$ (green). 
    A larger $r_{\rm ext}$ results in a steeper power-law slope. 
    \textbf{Middle:} Distribution of the power-law index of the density profile $\gamma$, measured at $r_{\rm ext} = 1.5\epsilon_g$ (purple) and $r_{\rm ext} = 2\epsilon_g$ (green). 
    For $r_{\rm ext} = 1.5\epsilon_g$, the distribution peaks at $\gamma=1.4$, while for $r_{\rm ext} = 2.0\epsilon_g$, the distribution peaks at $\gamma = 1.9$. 
    We plot the power-index estimate in \citet{Kelley2017b} for comparison.
    \textbf{Right:} Distribution of density extrapolated to $10\,{\rm pc}$. We compare the two $r_{\rm ext}$ values. The extrapolated density is sensitive to the change in $r_{\rm ext}$: $r_{\rm ext} = 1.5\epsilon_g$ gives a distribution centered at $10\, M_\odot/{\rm pc}^3$, while $r_{\rm ext} = 2.0\epsilon_g$ gives a distribution centered at $100\, M_\odot/{\rm pc}^3$.}
    \label{fig:density}
\end{figure*}

In \astrid, we have already accounted for the dynamical friction timescale above the resolution limit, leading to significant delays of in-simulation mergers compared to the traditional MBH repositioning methods. 
However, dynamical friction will continue to dominate over other delay processes on scales of $10\sim 100$ pc \citep[e.g.][]{Kelley2017a}, which is beyond our current resolution. 
In this section, we will compute the unresolved DF timescales for the MBH mergers, and compare with the in-simulation DF timescale.

For the in-simulation DF time, we measure it in the following way: for each black hole pair that merge within the simulation at $z_{\rm merge}$, we track their trajectories before the merger event, and find the redshift $z_{\rm encounter}$ at which they are first within $2\epsilon_g$ of each other. 
$z_{\rm encounter}$ is the approximate time at which the BHs would merge if we did not account for the dynamical friction time at all (note that under the reposition model, BHs usually merge even before $z_{\rm encounter}$). 
We consider the time difference between $z_{\rm encounter}$ and $z_{\rm merge}$ as the in-simulation DF time, $T_{\rm DF, sim}$. 
Among all BH mergers in the simulation, 5713 mergers ($\sim 1.4\%$ of the whole merger population) happen at the first encounter.

For the post-processed DF time $T_{\rm DF, post}$, we adopt the treatment from \citet{Merritt2013} and \cite{Dosopoulou2017}, who modifies the Chandrasekhar formalism \citep[e.g.][]{Chandrasekhar1943,Binney2008} to include the effect of the secondary BH embedded in a tight core of stars brought in from the secondary galaxy. 
The increased dynamical friction allows the secondary to sink faster towards the primary galaxy’s center, and thus the resulting dynamical friction time is less than the prediction from the canonical \cite{Binney2008} treatment assuming a bare BH. 
In \citet{Dosopoulou2017} the assumption is that the mass of stars bound to the secondary BH is 1000 times the mass of the BH itself, and the resulting dynamical friction timescale is:

\begin{equation}
\label{eq:tdf}
    t_{\rm DF,post} = 0.12\, {\rm Gyr} \left(\frac{r}{10{\rm kpc}}\right)^2 \left(\frac{\sigma}{300{\rm km/s}}\right) \left(\frac{10^8 M_\odot}{M_{\rm 2}}\right) \frac{1}{{\rm log}(\Lambda)},
\end{equation}
where $\text{log}(\Lambda)$ is the Coulomb logarithm, $M_2$ is the mass of the secondary black hole. 
For the initial separation $r$, we use the radius of the circular orbit that has the same energy as the last orbit of the binary before the (numerical) merger. 
Note that the model in Equation \ref{eq:tdf} does not account for the effect of non-circular orbits on the DF time. 
Taking the eccentricity into consideration can further reduce the estimated DF time \citep[e.g.][]{Taffoni2003}. 
Following the method in \cite[][]{Chen2021},we compute the Coulomb logarithm by:
\begin{equation}
    \Lambda = \frac{b_{\rm max}}{(GM_{\rm 2})/v_{\rm BH}^2}, \; b_{\rm max} = 10\text{ ckpc}/h,
\end{equation}
where $M_{\rm 2}$ is the mass of the secondary black hole and $v_{\rm BH}$ is the velocity of the secondary black hole with respect to the host galaxy center.

Figure \ref{fig:t_df} shows the comparison between the in-simulation dynamical friction time and the post-processed dynamical friction time from above. 
The top panel shows the overall distributions of the DF times. 
The two distributions are on the same order of magnitude at around $10^2$ Myrs, with a range from 10 Myrs to 1 Gyrs. 
For most BH pairs, $T_{\rm DF, sim}$ is longer than $T_{\rm DF, post}$. 
This means that accounting for dynamical friction to the simulation, we have already included about half of the total dynamical friction delay effects. 
Note that both DF timescales are shorter than 1 Gyr. In the case of the resolved DF time, this is mainly due to the fact that most of the black holes have not existed for more than 1 Gyr at $z=3$.

In the bottom panel of Figure \ref{fig:t_df}, we show the relation between the two DF times and the binary mass ratio $q$. 
Here we only show the times involving at least one non-seed MBH, defined as mergers with $M_1>2M_{1,\rm seed}$. 
This is because our merger population is dominated by seed MBHs which have not grown out of their dynamical mass and thus the in-simulation DF time estimation is not exact. 
From Equation \ref{eq:tdf}, we can see that the DF time is correlated with the mass of the primary galaxy (and MBH) through $\sigma$, and that it is inversely proportional to the secondary black hole mass. 
Hence, we expect that minor mergers will have longer decay timescales, and in the plot we do see a negative correlation between $T_{\rm DF,post}$ and $q$. 
For the in-simulation DF, although this relation is not imposed explicitly, we still observe a negative correlation between $q$ and the DF time. 
This indicates that the negative correlation is still captured by the in-simulation dynamical friction modeling.

\subsection{Loss Cone Scattering and Gravitational Wave Hardening}
\label{lcgw}
\begin{figure*}
\centering
    \includegraphics[width=0.98
\textwidth]{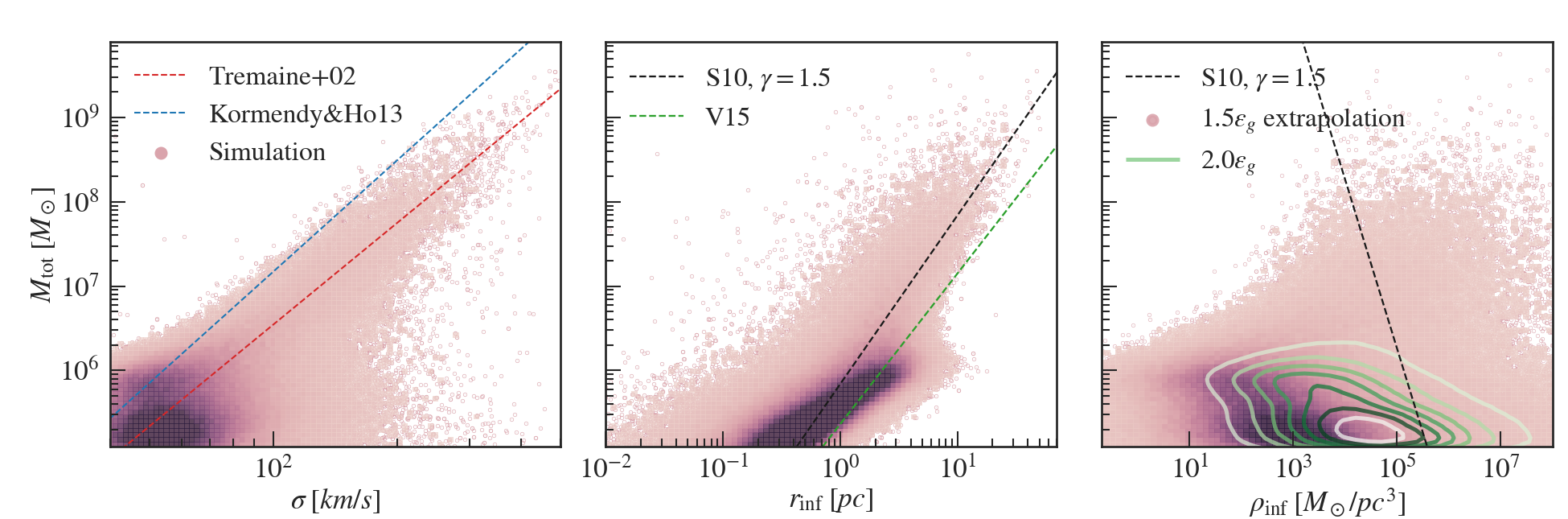}
    \caption{Variables used to calculate dynamical friction and binary hardening timescales. 
    \textbf{Left:} $M_{\rm tot}-\sigma$ relation measured from all the binaries at the time of merger in the simulation, compared to the analytical relation given in \citet{Tremaine2002} and \citet{Kormendy2013}. \textbf{Middle:} The influence radius derived from $\gamma$ and $\sigma$ measured the simulation, compared with the analytical model used in \citet{Vasiliev2015} (green dashed line), and in \citet{Sesana2010} with $\gamma=1.5$ (black dashed line). 
    Our measured $\sigma$ and $r_{\rm inf}$ are both close to the analytical models. 
    \textbf{Right:} Density at influence radius extrapolated from the simulation. 
    To illustrate the effect of extrapolation scales on $\rho_{\rm inf}$, we show the resulting extrapolation from both $1.5 \epsilon_g$ (pink dots) and $2.0 \epsilon_g$ (green contour). As was demonstrated in Figure~\ref{fig:density}, the density extrapolation is sensitive to the starting point of the extrapolation.  However, even the extrapolated density from an outer radius is smaller compared with the analytical model used in \citet{Sesana2010} with $\gamma=1.5$ (black dashed line).}
    \label{fig:vars}
\end{figure*}

Once the two MBHs become gravitationally bound, the dynamical friction formalism is no longer a valid approximation, and individual interactions between singular stars and the binary must be considered. 
These interactions extract angular momentum from the binary, driving them closer to each other \citep[e.g.][]{Merritt2013,Vasiliev2013}. 
This regime is the loss-cone scattering (LC) regime, which refers to the specific cone in parameter space where stars have to exist in order to extract angular momentum from the binary \citep[e.g.][]{Frank1976,Lightman1977}. 
On even smaller scales, the binary will enter the gravitational wave regime where it will evolve until coalescence. 
Once the binary enters the gravitational wave regime, its dynamics follow the formalism of \citet{Peters1964} at small separations of $10-1000$ Schwarzschild radii.

For the loss-cone scattering and gravitational-wave hardening phase, we adopt the analytical prescription in \cite{Vasiliev2015} (V15 hereafter). 
However, the time estimation in Equation (25) of V15 assumes a single family of $M_{\rm tot}-\sigma_{\rm inf}-r_{\rm inf}$ relation, and thus it may over-simplify the properties of the galaxies hosting the merger events. 
Hence we will adopt the V15 formalism but with some slight changes, so that we can use the host galaxy properties measured from the simulation. 
In this Section, we first explain how we measure the relevant galaxy properties, then we give our binary hardening time estimation by combining analytic modeling with the measured properties.

\subsubsection{Extrapolated Galaxy Properties}
\label{subsec:properties}
To compute the hardening time for the binaries, the important quantities to measure are: the influence radius $r_{\rm inf}$ defined as the radius containing a stellar mass equal to two times the binary mass, the velocity dispersion of stars at the influence radius $\sigma_{\rm inf}$, the power-law slope of the stellar density profile $\gamma$, and the stellar density at the influence radius $\rho_{\rm inf}$. 
Since the binary hardening phase begins after the dynamical friction phase, we use the snapshot immediately following the numerical merger to measure these properties. 

Among the quantities above, the velocity dispersion can be measured directly from the simulation without extrapolation (for an isothermal sphere, the velocity dispersion is independent of radius). 
Therefore, we take the approximation that $\sigma_{\rm inf} = \sigma_{\rm gal}$, and use the measured velocity dispersion within the half-mass radius of the host galaxy.

The next galaxy property we measure from the simulation is the power-law slope $\gamma$ of the stellar density profile. 
In Figure \ref{fig:prof}, we show three examples of the density profiles of dark matter, gas and stars for galaxies hosting recently merged binaries. 
We show the profiles of a massive binary with $M_{\rm tot}=5.6\times 10^8 M_\odot$, a less massive one with $M_{\rm tot}=7.6\times 10^6 M_\odot$ at $z=3$, and a seed-mass binary with $M_{\rm tot} = 1.9\times 10^6M_\odot$.
For the most massive binary (top), the stellar density is the dominant component on scales below $\sim 10$ ckpc/$h$. 
The stellar density profile follows a power law down to the gravitational softening length $\epsilon_g$, where the profile flattens due to gravitational softening. 
In the case of the medium-mass binary, the density of all three components is comparable at $r<10$ ckpc$/h$, and the density profile flattens at a larger radius compared to the massive one.
In the third case of a seed-seed merger, the mass of the host galaxy is high relative to the binary mass.
The binary is not the most massive MBHs in this galaxy, but the merger still occurs at a relatively central region.
We note that this binary belongs to the seed MBH population that still merge after the post-processing delays.

As we do not resolve the scale of interest for the loss-cone scattering, we assume that below a scale $r_{\rm ext}$ close to the resolution limit $\epsilon_g$, the stellar density profile follows a single power law $\rho \propto r^{-\gamma}$.
By doing so, we are able to extrapolate the stellar density to the inner region of the host galaxy. 
To measure the value of $\gamma$, we take the measured density from 10 bins just above $r_{\rm ext}$, and fit it to the power law profile. 
Our choice of $r_{\rm ext}$ is motivated by the flattening of the profile at $\sim 1.5\epsilon_g$ in Figure \ref{fig:prof} and the fact that gravity is not well-resolved within $\sim 2\epsilon_g$. 
Since the exact scale on which the simulation density becomes unrealistic is uncertain, we use both $r_{\rm ext}=1.5\epsilon_g$ and $r_{\rm ext}=2.0\epsilon_g$ to bracket our predictions.

The left panel of Figure \ref{fig:density} shows the measured stellar density profiles and extrapolations beyond $r_{\rm ext}$ for all binaries in \astrid. 
We show the median as well as the 95\% contour of the measured density, and we compare the power-law extrapolation from $r_{\rm ext}=1.5\epsilon_g$ and $r_{\rm ext}=2.0\epsilon_g$. 
From the comparison, we see that the measurement of $\gamma$ is sensitive to the extrapolation scale, and that larger $r_{\rm ext}$ results in a steeper power-law slope and thus a higher density at the inner region. 
However, we also note that the shift due to $r_{\rm ext}$ is comparable to the width of the distribution, and that both measurements are consistent with the values assumed in various binary hardening models.

This is further illustrated by the middle/right panel of Figure \ref{fig:density}, where we show the distribution of the measured $\gamma$ and the density extrapolated to $10$ pc. 
For $r_{\rm ext} = 1.5\epsilon_g$, the distribution peaks at $\gamma=1.4$, while for $r_{\rm ext} = 2.0\epsilon_g$, the distribution peaks at $\gamma = 1.9$. 
These values are consistent with the range of values used in most loss-cone scattering models \citep[e.g.][]{Sesana2010,Merritt2013,Vasiliev2015,Sesana2015}. 
In the figure, we also compared our distributions with the measured distribution in \citet{Kelley2017b} from the Illustris simulation. 
Compared to \citet{Kelley2017b}, our measured profiles are significantly steeper, which also lead to a higher extrapolated density at $r=10pc$.
Our simulation has a higher resolution than Illustris, and thus resolves the stellar density profiles better on kpc scales.
This is also due to the fact that we begin our extrapolation at different scales: \citet{Kelley2017b} uses the inner-most eight density bins that contains at least four particles, which could lie well below the gravitational softening. 
From the right panel, we see that the extrapolated density is sensitive to the change in $r_{\rm ext}$: $r_{\rm ext} = 1.5\epsilon_g$ gives a distribution centered at $10\, M_\odot/pc^3$, while $r_{\rm ext} = 2.0\epsilon_g$ gives a distribution centered at $100\, M_\odot/pc^3$. 
The order-of-magnitude difference motivates us to propagate the uncertainty in $r_{\rm ext}$ throughout subsequent analyses, as it may have non-trivial impacts on the final merger rate predictions from the simulation.

Finally, we compute $r_{\rm inf}$ and $\rho_{\rm inf}$ from the quantities measured above. 
As we cannot resolve the inner cusp of the galaxies in our simulation, a direct measurement of $r_{\rm inf}$ is not possible. 
To estimate the influence radius, we adopt the analytical relation \citep[e.g.][]{Sesana2010}:
\begin{equation}
    r_{\rm inf} = (3-\gamma)\frac{GM_{\rm tot}}{\sigma_{\rm inf}^2},
\end{equation}
where $\gamma$ is the density power law slope we just showed, and $\sigma_{\rm inf}$ is approximated by the measured galaxy velocity dispersion.

To get the density at the influence radius $\rho_{\rm inf}$, we extrapolate the power-law relation of the density profile down to $r_{\rm inf}$, using the measured $\gamma$ and $\rho$. 
Note that our simulation does not resolve the high-density peaks below our resolution, or nuclear star clusters, and thus the extrapolated $\rho_{\rm inf}$ is likely a lower limit. 
Moreover, since the nuclear star clusters are not resolved, we do not account for effects such as tidal disruption, which can to a shorter binary hardening time \citep[e.g.][]{Arca-Sedda2018,Biava2019,Ogiya2020}. 

Figure \ref{fig:vars} shows all of the measured or derived variables for computing the binary hardening timescales, and their relation with the binary mass. 
The $M_{\rm BH}-\sigma$ relation follows the relation in \citet{Tremaine2002} for binaries with $M_{\rm tot}>2\times 10^6 M_\odot$, but is flatter compared to the relation in \citet{Kormendy2013}. 
There is a large scatter in $\sigma$ for seed-mass binaries. 
Since the influence radius $r_{\rm inf}$ is derived from $\sigma$, $\gamma$ and the binary mass, we expect it to stay close to the analytical models from binary hardening papers. 
Here we compared it with the analytical model adopted in \citet{Sesana2010} and \citet{Vasiliev2015}. 
Our values are in line with the \citet{Sesana2010} model with a constant $\gamma=1.5$, although the scatter is large. 
This is also consistent with the fact that our distribution in $\gamma$ peaks around $\gamma = 1.4$ when measured at $r_{\rm ext} = 1.5\epsilon_g$.  

Finally, in the right panel we show the density at the influence radius extrapolated from the simulation. 
To illustrate the effect of extrapolation scales on $\rho_{\rm inf}$, we show the resulting extrapolation from both $1.5 \epsilon_g$ and $2.0 \epsilon_g$. 
As shown in Figure,\ref{fig:density}, the density extrapolation is very sensitive to the starting point of the extrapolation. 
Shifting the starting point by $0.5 \epsilon_g =0.75 {\rm ckpc}/h$ can result in an order of magnitude difference in $\rho_{\rm inf}$. 
However, we note that even the density extrapolated from the outer radius is smaller than the analytical model used in \citet{Sesana2010} with $\gamma=1.5$.

\subsubsection{Binary Hardening Timescales}
\begin{figure}
    \includegraphics[width=0.49\textwidth]{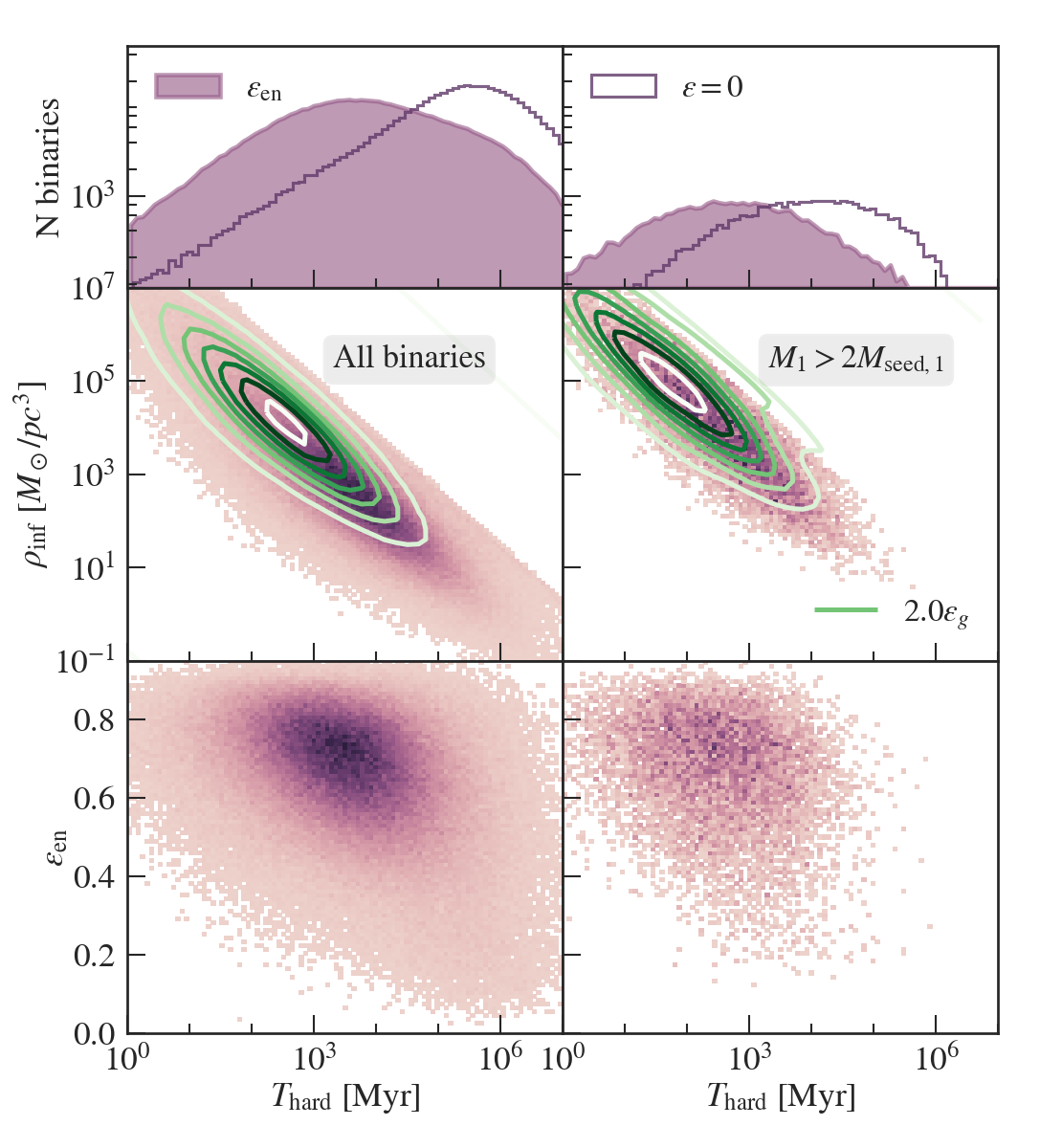}
    \caption{\textbf{Top:} The distribution of the loss-cone and gravitational-wave hardening time for all binaries in the simulation. Here we use $r_{\rm ext}$ from $1.5\epsilon_g$. The shaded distribution is computed using the measured eccentricity $\epsilon_{\rm en}$. If we assume $\epsilon=0$ (unshaded), the decay timescales will generally be longer by a factor of $\sim 100$ and peak at 100 Gyr, which is much longer than a Hubble time.
    \textbf{Middle:} the relation between the hardening timescale and the density at influence radius $\rho_{\rm inf}$. The timescale is negatively correlated with $\rho_{\rm inf}$. Changing $r_{\rm ext}$ from $1.5\epsilon_g$ (pink dots) to $2.0 \epsilon_g$ (green contours) shortens the hardening timescale. The right panel shows a clearer dependency when we remove the seed population. 
    \textbf{Bottom:} the relation between the hardening timescale and the measured eccentricity. We see a weak negative correlation between $T_{\rm hard}$ and $\epsilon_{\rm en}$.}
    \label{fig:tlc_s15_v15}
\end{figure}

After measuring the quantities of interested for computing the binary hardening time, we will proceed to describe the analytical model for estimating the hardening timescale. 
As was mentioned earlier, we base most of our model on \citet{Vasiliev2015} (V15), with appropriate changes to incorporate information from the simulation.

V15 models a separation-dependent LC hardening rate by:
\begin{equation}
    S_{*}(a) = \mu S_{\rm inf} \left( \frac{a}{a_h} \right)^\nu,
\end{equation}
where $a$ is the binary separation, $a_h$ is the hardening radius given by 
$a_h = \frac{q}{4(1+q)^2}r_{\rm inf}$, 
$\mu$ is the filling fraction of the loss cone, and $\nu$ characterizes the radial dependence of the hardening rate. 
We adopt the fiducial values of $\mu=0.3$ and $\nu= 0.4$ from V15. 
$S_{\rm inf}$ is the full LC hardening rate at the influence radius given by:
\begin{equation}
  S_{\rm inf} = H \frac{G \rho_{\rm inf}}{\sigma_{\rm inf}},
\end{equation}
where $\sigma_{\rm inf}$ and $\rho_{\rm inf}$ are the velocity dispersion and stellar density at the influence radius $r_{\rm inf}$,
and $H$ is a constant LC hardening rate given by $H=2\pi A$, with $A=4$ in V15. 
This value is slightly larger than the $H=15$ rate given by \citet{Sesana2015}.

At a closer separation, GW emission becomes the dominant channel for binary energy loss. 
The hardening rate in the GW regime is given by \citep[][]{Peters1964}:
\begin{equation}
    S_{\rm GW}(a) = \frac{1}{a^5} \frac{64G^3M_1M_2M_{\rm tot}F(\epsilon)}{5c^2},
\end{equation}
where $\epsilon$ is the eccentricity of the binary orbit and 
\begin{equation}
\label{eq:fe}
    F(\epsilon) = (1-\epsilon^2)^{-7/2}[1+(73/24)\epsilon^2+(37/96)\epsilon^4]
\end{equation}
accounts for the eccentricity dependence of the GW hardening rate.

The separation at which the binary spends the most time, $a_{\rm gw}$, is calculated by setting $S_{*}(a)=S_{\rm GW}(a)$, which leads to:
\begin{equation}
    a_{\rm GW} = \left(  \frac{64G^3M_1M_2M_{\rm tot}F(\epsilon)}{5c^2} \frac{a_{h}^{\nu}\sigma_{\rm inf}}{\mu S_{\rm inf}} \right)^{1/(5+\nu)}
\end{equation}
Finally, we can estimate the LC+GW hardening timescale by:
\begin{equation}
\label{eq:thard_gw}
    T_{\rm hard}^{\epsilon_{gw}} = \frac{1}{S_{*}(a_{\rm GW}) \times a_{\rm GW}}.
\end{equation}
Note that in this expression, we have only accounted for the eccentricity dependence during the GW hardening stage, and thus the superscript $\epsilon_{\rm gw}$. 
However, the orbital eccentricity also evolves during the LC scattering phase and can impact the hardening time. V15 models this effect by:
\begin{equation}
\label{eq:thard_final}
    T_{\rm hard} = T_{\rm hard}^{\epsilon_{gw} = 0} \times
    (1-\epsilon^2)[k+(1-k)(1-\epsilon^2)^4]
\end{equation}
where $k=0.4+0.1\, {\rm log}_{10}(M_{\rm tot}/10^8\, M_\odot)$.
At higher eccentricities, Equation \ref{eq:thard_gw} and \ref{eq:thard_final} give similar results, but for $\epsilon\sim 0$, the former underestimates the hardening timescale by a factor of $\sim 3$.

\begin{figure*}
    \includegraphics[width=0.85\textwidth]{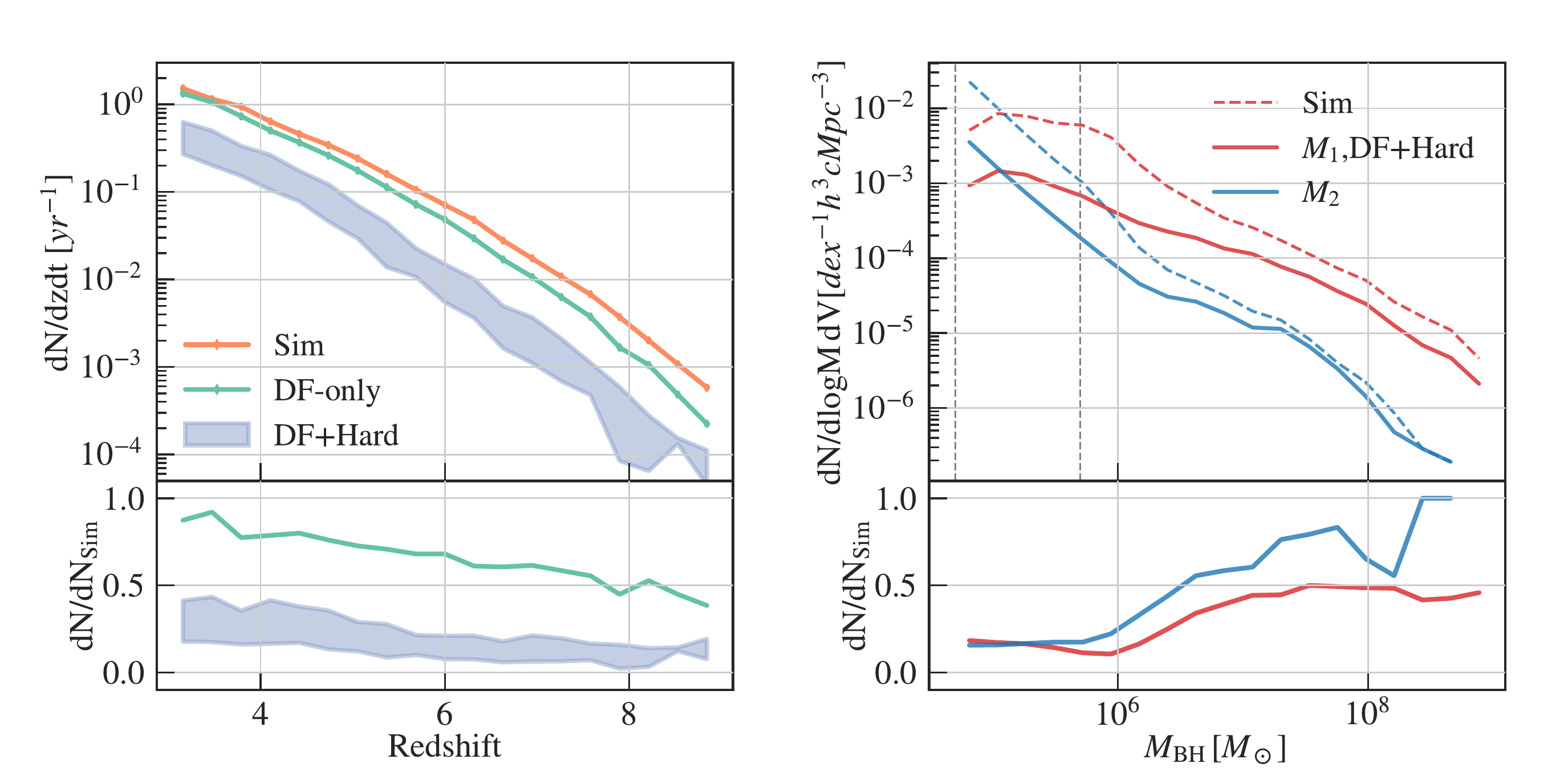}
    \caption{\textbf{Left:} The merger rates for all binary population in \astrid~down to $z=3$ with different levels of delays. Without considering any post-processing delays (\textbf{orange}), we expect a total of $\sim 2$ mergers per year of observation down to $z=3$. 
    The rate when considering only the DF delay (\textbf{green}) has an at most $50\%$ decrease compared to the raw rate at the highest redshifts. 
    The binary hardening time has the most significant effect in reducing the merger rate (\textbf{purple}).
    The upper limit of the band assumes $r_{\rm ext}=2\epsilon_g$, and the lower limit assumes $r_{\rm ext}=1.5\epsilon_g$.
    The bottom panel shows the ratio between the delayed merger rates and the simulation merger rates.
    \textbf{Left:} The mass distribution of the two MBHs involved in the mergers. The red curves correspond to the more massive MBH and the blue curves correspond to the less massive MBH. 
    The mass distribution of the simulation mergers is plotted in dashed lines, and that of the delayed mergers is plotted in solid lines. 
    The bottom panel shows the ratio between the mass distributions of simulation mergers and delayed mergers.
    The seed-mass mergers (enclosed in the vertical dashed lines) are suppressed most strongly by a factor of $\sim 6$.
   }
    \label{fig:rates_single}
\end{figure*}

For the binaries in our simulation, we use the galaxy and binary properties shown in Section \ref{subsec:properties}, together with the above formalism to estimate the binary hardening time. 
Note that the hardening timescale depends on the orbital eccentricity as the BHs enter the hardening regime: more eccentric orbits merge faster compare to circular ones. 
To take this effect into account, we use the orbital eccentricity shown in Section \ref{sec:ecc} as a proxy for the orbital eccentricity at the beginning of the binary hardening phase, assuming that the post-processed dynamical friction does not change the orbital eccentricity greatly \citep[e.g.][]{Colpi1999,Hashimoto2003}. 
However, some studies also suggested that this might not be the case, as the rotation of gas on $10^2$ pc scales can affect the BH angular momentum and eccentricity \citep[e.g.][]{Dotti2007,Bonetti2020}. 
We also note that the  galaxy properties we put into the calculation are instantaneous properties from the simulation after the BHs go through the numerical merger. 
Given that the galaxy and central stellar densities will only grow with time (as well as the BH masses), our estimations are likely upper limits of the hardening time.

Figure \ref{fig:tlc_s15_v15} shows the relation between the binary hardening time and $\rho_{\rm inf}$ as well as the energy-based eccentricity $\epsilon_{\rm en}$. 
We also show the 1D distribution of hardening times. 
The left column includes all binaries in the catalog, while the right column only includes binaries with $M_1>2M_{\rm seed,1}$.
For all binaries, given our measured initial eccentricities, the hardening timescale falls between 100 Myrs and 100 Gyrs, with a peak around 5 Gyrs. 
The timescale is strongly correlated with $\rho_{\rm inf}$ and therefore also $r_{\rm ext}$. 
Changing the value of $r_{\rm ext}$ from $1.5\epsilon_g$ to $2.0 \epsilon_g$ leads to a shorter estimated hardening timescale.
This is because higher stellar density leads to shorter hardening timescales, as the LC stars can more efficiently carry away the energy from the binary. 
In fact, we find that the inner stellar density is the most important property for determining the hardening timescale. 
Nonetheless, in both cases the hardening timescale is much longer than the dynamical friction timescale. 
Note that if we do not account for the eccentricities of the binary orbits, the decay timescales will generally be longer by a factor of $\sim 100$ and peak at 100 Gyr, which is much longer than a Hubble time.

The bottom row shows the relation between the hardening timescale and the measured eccentricity. When looking at the whole binary population, we see a negative correlation between $T_{\rm hard}$ and $\epsilon_{\rm en}$. 
This is expected as eccentric orbits have accelerated hardening rates. 
However, when we only focus on the non-seed mergers, the $\epsilon_{\rm en}$ dependence is washed out by the strong correlation with $\rho_{\rm inf}$.

Because of the strong dependence of the delay timescale on the uncertain variable $\rho_{\rm inf}$, we will propagate this uncertainty to the merger rate predictions in the next session, and characterize how the uncertainty due to numerical resolution affects the mergers in \astrid.
\section{MBH merger rate and Host galaxy properties}
\label{sec:rate}
After characterizing the delay time, in this section we present the rate at which GW signals from MBH mergers will reach the earth, taking into account the sub-resolution delay processes.
We also examine how the DF and binary hardening delay affects different population of MBH mergers.
Finally, we investigate the galaxy properties for different part of the merger population.

\subsection{Merger Rate Predictions}
\begin{figure*}
    \includegraphics[width=0.93\textwidth]{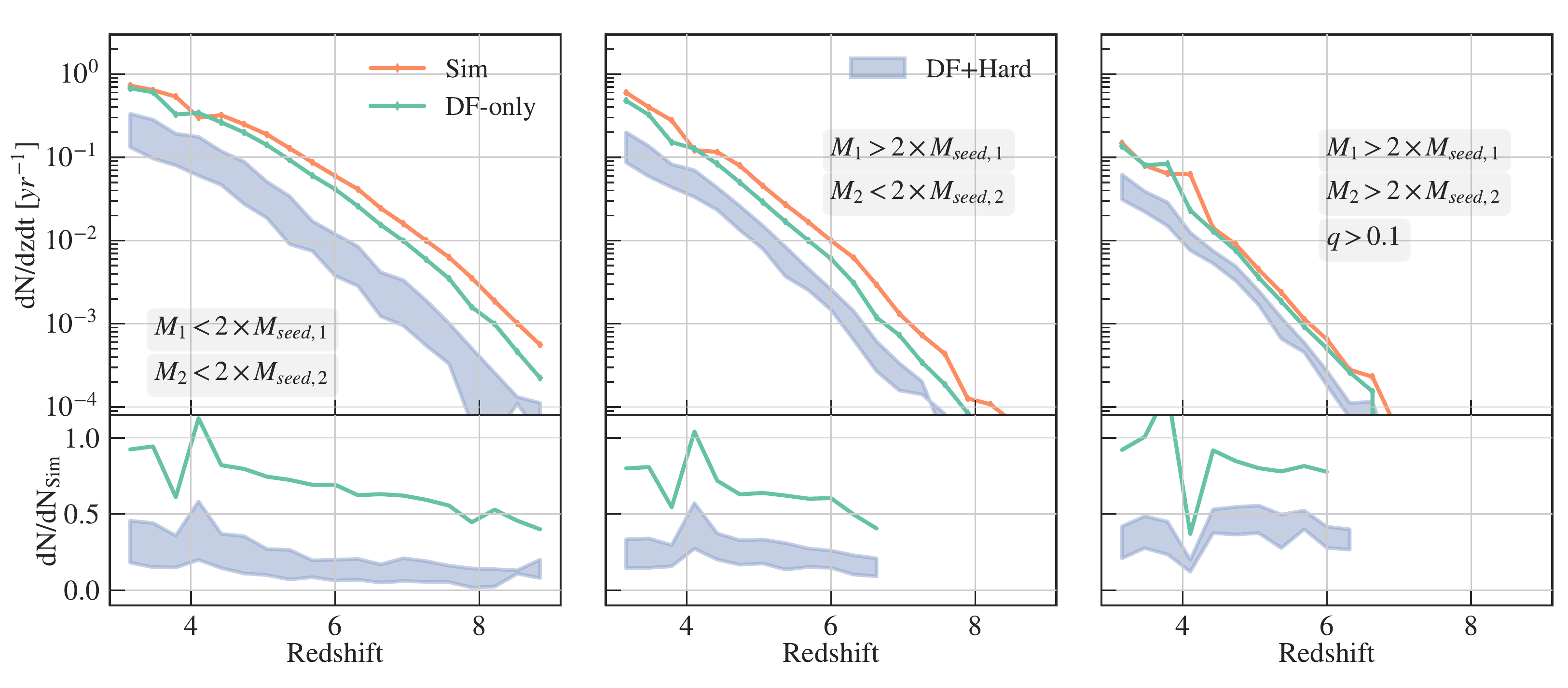}
    \caption{Merger rates for different mass cuts and mass-ratio cuts. 
    \textbf{Left:} The merger rates for the seed-mass population, where the masses of both MBHs are less than two times their seed masses.
    The colors are the same as in Figure \ref{fig:rates_single}.
    Compared to \ref{fig:rates_single}, this population makes up $\sim 60\%$ of the mergers.
    \textbf{Middle:} Merger rate for MBHs with only one of the two grown out of the seed mass. 
    This rate makes up $\sim 30\%$ of the entire merger population. 
    Compared to the seed-seed mergers, here we see less mergers at high redshifts, but a similar rate at $z=3$.
    \textbf{Right:} Mergers with both MBHs larger than two times their seed masses and with $q>0.1$. When constrained to major and non-seed mergers, the effect of DF is barely noticeable.
    The DF+Hard delayed rate makes up $50\%$ of the total rate.
    The lower panels show the ratio between the delayed merger rates and the simulation merger rates.}
    \label{fig:rates}
\end{figure*}

We calculate the rate by integrating the number of mergers in the simulation over redshifts, incorporating the cosmic volume at the given redshift:
\begin{equation}
    \frac{{\rm d}N}{{\rm d}z\,{\rm d}t} =  \frac{{\rm d}^2 n(z)}{{\rm d}z\,{\rm d}V_c} \frac{{\rm d}z}{{\rm d}t} \frac{{\rm d}V_c}{{\rm d}z} \frac{1}{1+z}\,
\end{equation}
where ${\rm d}V_c$ is the comoving volume element of the universe at a given redshift and $n(z)$ is the number of mergers at that redshift. The $1/(1+z)$ term redshifts the infinitesimal time element in ${\rm d}z/{\rm d}t$ to the observer frame time interval.

To calculate this rate from our simulation, we take the finite-interval approximation:
\begin{equation}
    \frac{{\rm d}^2 n(z)}{{\rm d}z\,{\rm d}V_c} = \frac{N(z)}{\Delta z\,V_{\rm sim}},
\end{equation}
where $\Delta z$ is the width of the redshift bin, $N(z)$ is the total number of mergers within that redshift bin, and $V_{\rm sim} = (250\, {\rm Mpc}/h)^3$ is the volume of our simulation in comoving units.

To clearly see the effect of each stage of the delay, we calculate three different rates. 
We first compute the "Sim" rate which uses the numerical merging time as the redshift of the merger \citep[also see][]{DeGraf:2021inprep}. 
Then we add the post-processed DF time to the numerical merger time to compute the "DF-only" rate. 
Finally, we further account for the binary hardening timescales and calculate the "DF+Hard" rate. 

In the left panel of Figure \ref{fig:rates_single}, we plot the merger rates with different levels of post-processed delays, for the whole merger population in \astrid.
First, we notice that the number of mergers keeps increasing with decreasing redshift for all three models. 
This is because we keep seeding BHs as structures form and grow, and that we have not reached the peak in seeding rate at $z=3$. 
Without considering any post-processed delays ("Sim"), we expect a total of $\sim 1.8$ mergers per year of observation down to $z=3$. 
The post-processed DF time does not significantly impact the total observed merger rate ("DF-only"), with a $\sim 50\%$ decrease at the highest redshift ($z \sim 8$).
The binary hardening time has the most significant impact on the merger rate at all redshifts ("DF+Hard"). 
We see that the merger rate is reduced by a factor of $3\sim 7$ after adding the delay from binary hardening. 
The resulting merger rate is $0.3\sim 0.7$ at $z>3$.
Here the upper limit is given by assuming $r_{\rm ext} = 2\epsilon_g$ and the lower limit is given by $r_{\rm ext} = 1.5\epsilon_g$.
On the bottom panel, we show the ratio between the delayed merger rate and the simulation merger rate as a function of redshift.
For both DF-only and DF+Hard delays, the fractional rates get higher at lower redshifts.
This is a result of the high-redshift mergers being pushed down to low redshifts.

On the right panel, we show the mass distribution of the two MBHs involved in each merger. 
The dashed lines corresponds to the simulation merger without any delays, and the solid lines shows the distribution of the merger population after the DF+hardening delays.
First, we can see that both before and after the delay, the merger population is dominated by seed-mass mergers (the ones enclosed by the vertical dashed lines), with $M_1$ evenly distributed across the seed masses and $M_2$ concentrated on the lower-mass end of the seeds.
It is also this seed-mass merger population that gets suppressed the most by the delay.
From the ratio between the mass functions shown in the bottom panel, we see that for the seed-mass mergers, only $\sim 15\%$ still merge at $z>3$ after the delays, whereas at the high-mass end this fraction increases to $50\%$.

In order to disentangle different merger populations, in Figure \ref{fig:rates} we further split the rate by how many seed MBHs are involved in the merger. 
The left panel shows the merger rates for the seed-mass population, where the masses of both MBHs are below two times their seed masses.
This population makes up $\sim 60\%$ of the mergers.
At $z>5$, the seed-seed mergers are strongly suppressed by the binary hardening delays because the stellar density is relatively low.
The middle panel shows the mergers with the only more massive MBH grown beyond two times its seed mass.
At $z=3$, the rate from this group is comparable to the rate from the seed-seed mergers. However, the number decreases more steeply as we go to higher redshifts.
Compared to the seed-seed mergers, this group has a higher mass ratio and thus a longer DF time.
The effect of the binary hardening delay, however, is smaller because of the higher-density in the remnant galaxy.
Finally, on the right panel, we show the more massive and major mergers. 
Compared to the previous two groups where at least one seed-mass MBH is involved in the merger, the mergers from this group is $\sim 6$ times lower.
The effect of delay is also the smallest. 
In particular, we noticed that the DF-only rate is very similar to the simulation rate.
Even for this group where the effect of delays is the smallest, the merger rate is still suppressed by $>50\%$ at each redshift compared to the simulation merger rate.

\begin{figure*}
    \includegraphics[width=0.99\textwidth]{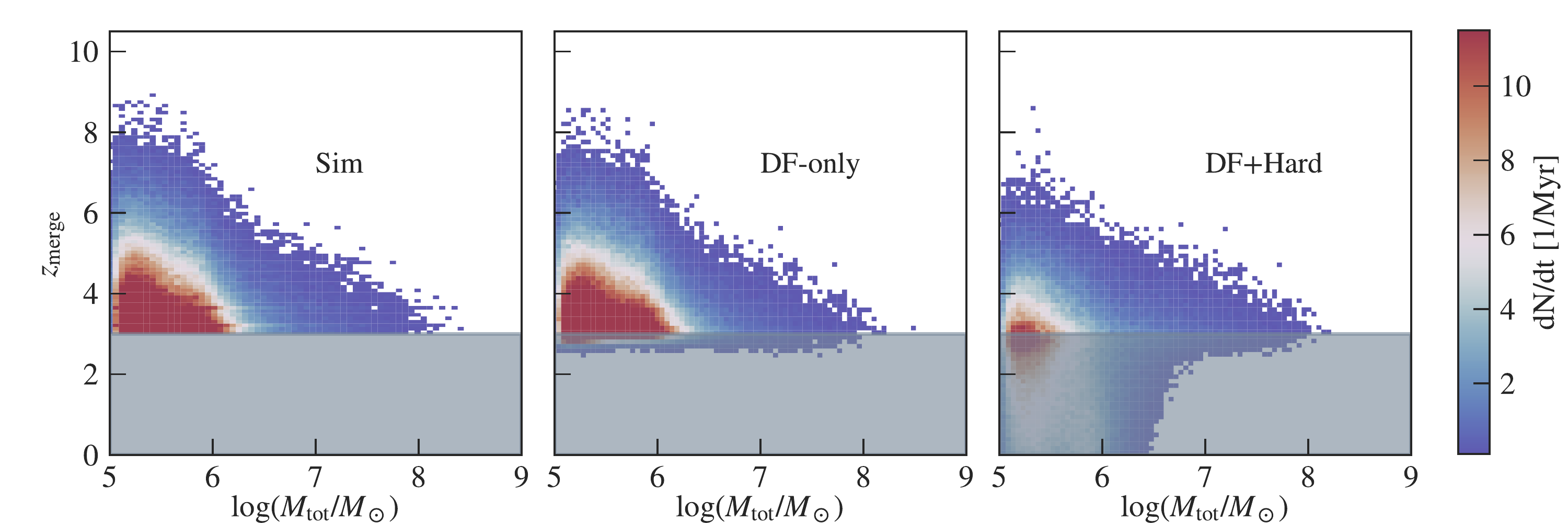}
    \caption{The distribution of mergers on the $M_{\rm tot}-z_{\rm merge}$ plane for the simulation and delayed mergers, color coded by the number of mergers per Myr. \textbf{Left:} The distribution for all mergers without delays.
    \textbf{Middle:} The same merger population with the post-merger DF time added. 
    Here, we see a slight shift of the merger population towards lower redshift, but nothing gets delayed below $z=2$. 
    \textbf{Right:} The distribution after considering both the DF delay and the hardening time. Note that since the latest redshift of the simulation is $z=3$, all the data points at $z<3$ (masked in grey) are results of delay from $z>3$ numerical mergers, and is not representative of all merger events at $z<3$. We see a significant shift of the mergers towards lower redshifts. The population most significantly shifted are the low-mass mergers with $M_{\rm tot}<10^{6.5}M_\odot$, while the most massive binaries are still able to merge at relatively high redshifts.}
    \label{fig:rates_mass}
\end{figure*}

Figure \ref{fig:rates_mass} shows the distribution of MBH mergers on the $M_{\rm tot}-z_{\rm merge}$ plane for both the simulation and delayed mergers, color coded by the number of mergers per Myr. Without any delay, the majority of the merger events are seed-seed mergers around $z=3-4$. As we would expect from the black hole mass growth over time, we see more massive mergers at lower redshifts. The middle panel shows the same merger population with the post-merger DF time added. As was discussed in the previous paragraph and in Section \ref{sec:df}, the post-processed DF peaks around $200$ Myrs and does not significantly delay the mergers. Here, we see a slight shift of the merger population towards lower redshift. 

On the right panel, we show the distribution after considering the DF delay and hardening phase. Note that since the final simulation output is at $z=3$, all the data points at $z<3$ are the results of delayed $z>3$ numerical mergers, and are not representative of all merger events at $z<3$. 
Compared with the other two panels, we see a significant shift of the mergers towards lower redshifts. 
The population that is most significantly shifted are the low-mass mergers with $M_{\rm tot}<10^{6.5}M_\odot$, while the most massive binaries are still able to merge at relatively high redshifts.
This is a consequence of the large hardening time scale of smaller BHs associated with lower $\rho_{\rm inf}$.

\subsection{Properties of High-z MBH Mergers}

From the previous section, we have seen that while some low-mass mergers are significantly delayed and do not merge at $z>3$, $\sim 15\%$ of them still do.
For the non-seed mergers, although the delay is generally less significant, we still see a $50\%$ decrease in merger rate when accounting for the delays.
Now we will investigate which part of the merger population gets significantly delayed, and which still manages to merge at high redshifts. 

\begin{figure*}
    \includegraphics[width=0.95\textwidth]{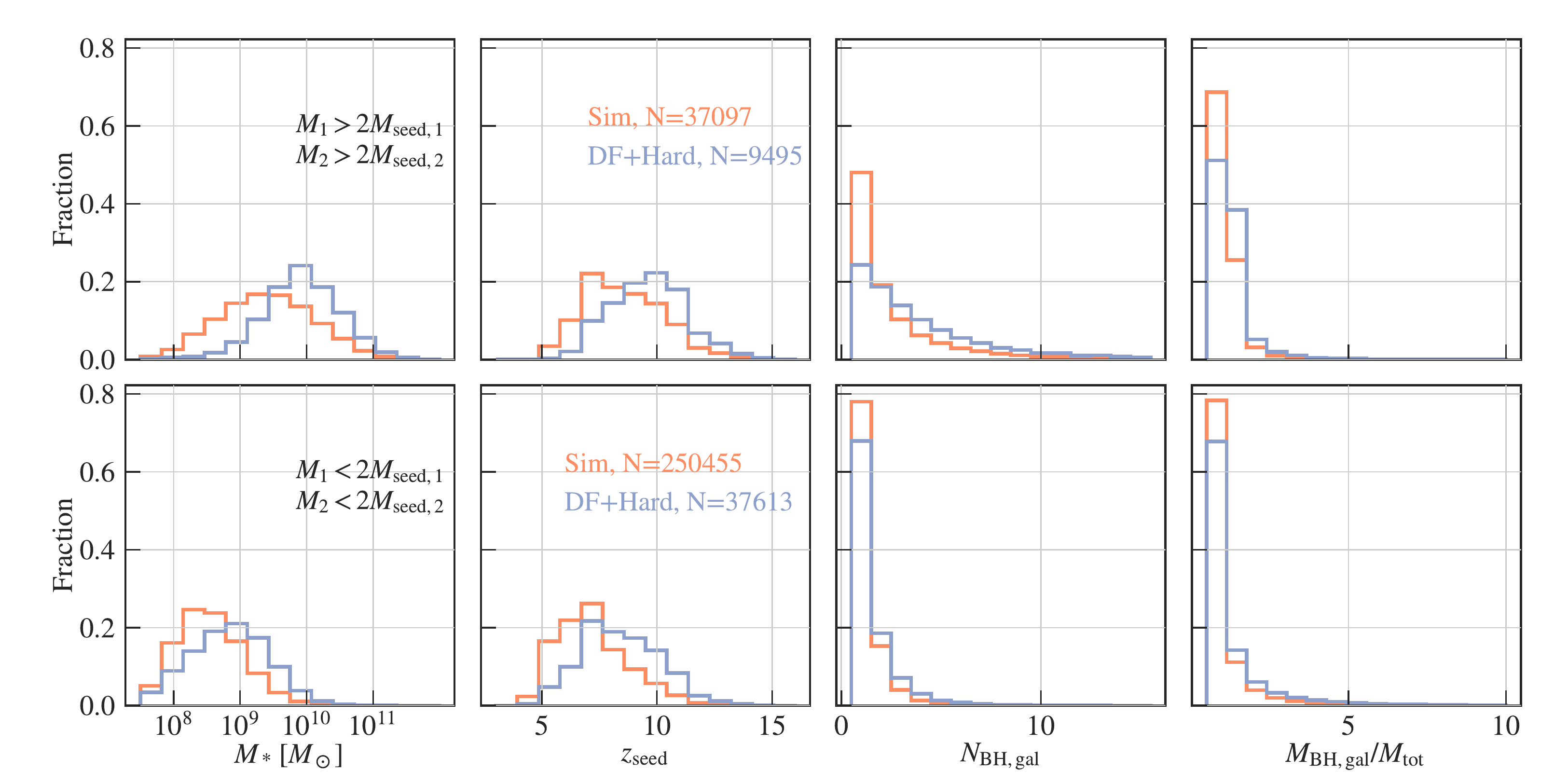}
    \caption{The fraction of the merger population in each bin of the galaxy stellar mass hosting the merger (\textbf{first column}), seeding redshifts of the merged MBHs (\textbf{second column}), number of MBHs in the host galaxy (\textbf{third column}), and the ratio between the total MBH mass in the host galaxy and the binary mass $M_{\rm tot}$ (\textbf{fourth column}). The top row shows the non-seed merger population, and the bottom row shows the seed-mass merger population. The simulation mergers are shown in orange and the DF+Hard delayed mergers are shown in purple. The total number of $z>3$ mergers in each population is shown in the second column with corresponding colors.}
    \label{fig:distributions}
\end{figure*}

In Figure \ref{fig:distributions} we show the properties of MBHs involved in both the simulation mergers and the delayed mergers. 
The top row shows the properties of the non-seed mergers, and the bottom row shows the properties of the seed-seed mergers.
We start by looking at the mass distribution of galaxies hosting the mergers (shown in the first column).
For the simulation mergers consisting of two non-seed MBHs, the masses of the host galaxies peak at $4\times 10^9 M_\odot$.
For systems that still merge after the delays, we see a clear shift towards the higher-end in stellar masses with a peak at $\sim 10^{10}M_\odot$.
This is because for more massive galaxies, the high stellar density enables more efficient hardening through loss-cone scattering, and thus the delay time is shorter (also see Figure \ref{fig:tlc_s15_v15}).
For mergers involving two MBH seeds shown on the bottom, we observe a similar trend.
Overall, seed mergers reside in less massive galaxies with stellar masses below $4\times 10^8 \, M_\odot$.
The delayed merger events also pick up the more massive galaxy population out of the simulation mergers with galaxy masses distributed around $10^9\,M_\odot$.

In addition to the stellar environment which plays an important role in the delay time estimation, the seeding redshift of the MBHs can also affect whether the two MBHs still merge at a high redshift after the delay.
This is shown in the second column of Figure \ref{fig:distributions}.
While the seeding redshift of the simulation merger MBHs is $z\sim 7$, the MBHs involved in delayed mergers are seeded as early as $z=10$.
For the seed-seed mergers shown on the bottom, the overall seeding redshift is lower, but we also see a shift towards higher redshift when comparing the delayed mergers to the simulation mergers.
The bias towards early MBH seeding for delayed mergers is also correlated with the higher host galaxy mass we have seen earlier: because the delayed mergers favor earlier seeds, they also tend to reside in galaxies that are massive enough at high redshifts to host an MBH seed.

On the right two columns, we examine the properties of other MBHs embedded in the host galaxy of the mergers.
The third column shows the total number of MBHs embedded in the host galaxy of the merger, in the snapshot immediately following the numerical merger (so the merging MBHs will be counted as one object).
The fourth column is the mass ratio between all MBHs in the host galaxy and the merging system.
For both the seed and non-seed merger populations, the merging system is the sole MBH in the host galaxy in a majority of mergers.
For the non-seed population, there is still a $>50\%$ fraction of mergers happening next to a third MBH (or more).
Interestingly, the delayed merger systems favor galaxies with more MBHs besides the merging ones (also correlated with larger galaxy masses).
Nonetheless, the merging system is still the most massive MBH in its host galaxy in the most cases when we look at the $M_{\rm BH,gal}/M_{\rm tot}$ ratio.

When constrained to seed-seed mergers, we see that $\sim 70\%$ of the mergers are the single MBH in the host galaxy.
The delayed mergers also tend to pick out the galaxies with more MBHs compared to the simulation mergers.
However, contrary to the non-seed case where the merging MBH is more massive than the other MBHs in the same galaxy, for seed-seed mergers that do occur near a third MBH, the mass of the third MBH is more likely to be larger.
This can be seen from the fact that the $N_{\rm BH, gal}$ distribution is more peaked at $N_{\rm BH, gal}>1$ compared to the $M_{\rm BH, gal}/M_{{\rm tot}}>1$ distribution (it means that if there is a third MBH, its mass can be larger than $M_{\rm tot}$ in some cases, resulting in the longer tail of $M_{\rm BH, gal}/M_{{\rm tot}}$).

From the investigations above, we conclude that the $z>3$ mergers after the DF and hardening delay make up a small and biased sample of the simulation mergers.
In particular, they are systems with MBHs seeded earlier and embedded in more massive galaxies compared to the overall simulation merger population.
Moreover, the majority of the merger remnant is the only MBH in its host galaxy, especially for the seed-mass mergers.
However, the delayed mergers tend to pick out more systems that have other nearby MBHs in the remnant galaxy compared to the overall simulation merger population.

\section{Gravitational Wave Emission from MBH Mergers}
\label{sec:snr}

With a catalog of merging binaries, their merging time, and orbital eccentricities, we can not only compute the rate of mergers reaching the Earth, but also predict the gravitational wave signal that can be observed from these sources. This section is dedicated to predicting the gravitational wave signal and detectability of the \astrid~mergers with LISA. We first briefly describe the characteristic strain for circular sources, and then we generalize to the signal from eccentric sources. After that, we combine with the LISA sensitivity curve and compute the signal-to-noise ratio (SNR) for each mergers in the simulation.

\subsection{Characteristic Strain of Circular Orbits}

MBH binaries provide a variety of signals measurable by LISA since their chirp evolution in the frequency domain occurs near the low-frequency band edge of the LISA sensitivity curve. 
Binaries with $10^5-10^7 M_\odot$ total mass will provide a measurable inspiral, merger, and ringdown leading to signals out to the cosmic horizon \citep{LISA2017arXiv170200786A}.
The binary inspiral is the initial stage of binary black hole coalescence when the two MBHs orbit one another at separations greater than the innermost stable circular orbit (ISCO; $R=6GM_{\rm BH}/c^2$). 
At these separations, the orbit is usually treated with a post-Newtonian formalism.
The merger stage follows the binary inspiral with a highly non-linear relativistic process. This process continues until the binary components  form a single event horizon, leading to ringdown.

We use the characteristic strain, $h_s$, to model the binary signal which accounts for the time the binary spends in each frequency bin \citep[][]{Finn2000}. 
The characteristic strain is given by \citep[e.g.][]{Moore2015}:
\begin{equation}
\label{eq:hc}
    h_s(f) = 4f^2 |\tilde{h}(f)|^2
\end{equation}
 where $\tilde{h}(f)$ represents the Fourier transform of a time domain signal.

 To generate the waveforms, we use the phenomenological waveform PhenomD \citep[][]{Husa2016,Khan2016} implemented within the \texttt{gwsnrcalc} Python package \citep[][]{Katz2019}. 
 The input parameters are the binary masses, merging redshift, and the dimensionless spins of the binary. 
 For the MBH masses, we do not account for mass growth after the numerical merger. 
 However, we note that the MBH can potentially gain a significant fraction of its mass during the $>1\,{\rm Gyr}$ of time in the dynamical friction \citep[e.g.][]{Banks2021} or loss-cone scattering phase. 
 The dimensionless spin $a$ characterizes the alignment of the spin angular momentum with the orbital angular momentum, and the value of $a$ ranges from $-1$ to $1$. 
 However, we do not have any information on the spin of the SMBHs in our simulation. 
 Therefore, following the argument in \citet{Katz2020}, we assume a constant dimensionless spin of $a_1=a_2=0.8$ for all binaries \citep[e.g.][]{Miller2007,Reynolds2013}.

 \begin{figure*}
\centering
    \includegraphics[width=0.93\textwidth]{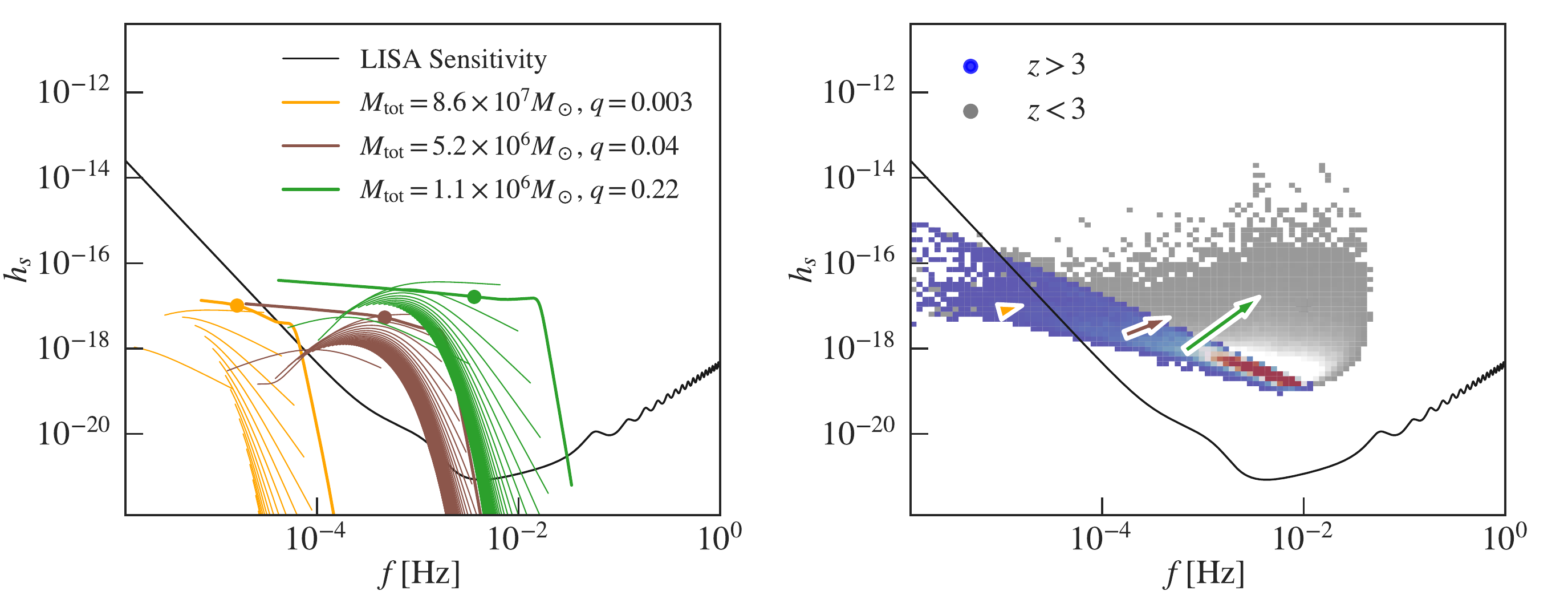}
    \caption{\textbf{Left:} Example waveforms for three binaries of different masses in \astrid. The thick curve shows the waveform assuming $\epsilon=0$, while the thin lines are the waveform assuming eccentric orbits. We also show the LISA sensitivity curve from \citet{LISA2017arXiv170200786A} (black solid) for comparison. The numerical merging time of all example binaries is $z\sim 3.1$.  \textbf{Right:} The $h-f$ distribution after applying the delay models. The arrows indicate the shifts in strain and frequency by the delay. Most signals are shifted to the upper-right due to the lower redshift of the merger after the delays. The light grey region shows the merger population delayed to $z<3$, which is not part of our prediction.}
    \label{fig:hc_fc}
\end{figure*}

In Figure \ref{fig:hc_fc}, we show the distribution of the merging frequency $f_{\rm merge}$ and the strain at this frequency for all binaries in the simulation, before and after applying the DF+Hardening delay models. 
To evaluate the detectability of the population with LISA, we also plot the proposed LISA sensitivity curves. 
We use the LISA sensitivity configuration from the LISA Mission Proposal \citep[][]{LISA2017arXiv170200786A}, and we use $h_N =\sqrt{S_N}$ \citep[][]{Moore2015} to convert from the proposed power spectral density $S_N$ to strain $h_N$. 

In the left panel, we show example waveforms for binaries of different masses but similar numerical merging time.
The thick curve shows the waveform assuming $\epsilon=0$, with the dot representing the merging frequency $f_{\rm merge}$. We will discuss the thin lines with non-zero eccentricities in later sections. From the example waveforms, we see that at a fixed source redshift, the more massive binary has a higher strain amplitude. 
However, this does not necessarily lead to a more significant detection, because the lower frequency at which the wave is emitted falls into the region where the LISA sensitivity is worse. 
Out of these three binaries, the two least massive binaries are detectable by LISA while the most massive one is not.
After the DF and hardening delays, all curves have higher strain amplitudes, as the strength of the signal is negatively correlated with redshift. 

After looking at individual cases, we turn to the whole binary population. 
On the right panel, we show the distribution of $f_{\rm merge}$ and $h_s(f_{\rm merge})$ for \astrid~mergers, after the post-processed delays.
We have masked the signals from $z<3$ mergers in light grey, as they are purely due to the post-processed delays, and are not part of our simulation predictions.
The majority of merger events within the simulation lie above the LISA sensitivity curve. 
From example waveforms, we see that once any given GW signal crosses the detector sensitivity curve, the ratio of the signal to the sensitivity curve rapidly increases by a few orders of magnitude. 
Since the merger population is dominated by seed-seed mergers, we see a peak around $f_{\rm merge}\sim 10^{-2}$Hz, corresponding to the example green curve. 
Finally, we demonstrate the shift of the signal due to the delay model by the colored arrows.
The tail of the arrows indicates the location of the frequency/strain before the delays.
The head of the arrows are the signals after the delays.
We see that in the example cases, the signal shifts to the high-strain, high-frequency region of the plane. 
This is mainly because of the delay of the mergers from $z>3$ to $z<3$.

\subsection{GW Signal from Eccentric Sources}
In the previous section we have shown a single $h_s-f$ relation by assuming circular orbits for the binaries. In this section, we will utilize the eccentricity measured from the simulation when calculating the strain and signal-to-noise ratio (SNR) for each binary. 

The GW strain from an individual, eccentric source can be related to that of a circular source as \citep[e.g.][]{Amaro-Seoane2010,Kelley2017b}:
\begin{equation}
\label{eq:ecc_hf}
    h_s^2(f_{\rm r}) = \left(\frac{2}n\right)^2 \sum_{n=1}^{\infty} h_{\rm r,circ}^2 (f_{\rm h}) g(n,\epsilon) |_{f_{\rm h} = f_{\rm r}/n},
\end{equation}
where $h_{\rm r,circ}$ is the characteristic strain of a circular source given by Equation \ref{eq:hc}, $g(n,\epsilon)$ is the GW frequency distribution function given by Equation 20 in \citet{Peters1963} with $\sum_{n=1}^{\infty} g(n,\epsilon) = F(\epsilon)$, where $F(\epsilon)$ is defined by Equation \ref{eq:fe}.

During the GW-driven inspiral, the orbital eccentricity also evolves according to \citet{Peters1964} Equation (5.7), such that it decays towards zero as the binary inspirals toward merger. This will affect the orbital frequency by:
\begin{equation}
\frac{f_{\rm orb}}{f_0} = \left[\frac{1-\epsilon_0^2}{1-\epsilon^2} \left(\frac{\epsilon}{\epsilon_0}\right)^{12/19}\left(\frac{1+\frac{121}{304}\epsilon^2}{1+\frac{121}{304}\epsilon_0^2}\right)^{870/2299}\right]^{-3/2} \ ,
\end{equation}
where $\epsilon_0$ is the initial eccentricity at the reference frequency $f_0$. 

In Figure \ref{fig:hc_fc}, the multiple thin lines are the waveforms  from higher-order harmonics assuming eccentric orbits. 
For circular orbits, the GW is emitted at a single frequency at a fixed separation, while the eccentric binaries emit GW at higher-order harmonics at a given time. 
One consequence of this is that the energy dissipated in higher-order harmonics is below the detection sensitivity, and thus the signal will be smaller compared with the circular orbits.

We note that when using the simulation measurement of the orbital eccentricity as the initial eccentricity in the inspiral phase, we did not account for the possible increase in $\epsilon$ during the loss-cone scattering phase \citep[see, e.g.][]{Sesana2010,Kelley2017b}.

\subsection{Detectability Prediction}

\begin{figure*}
\centering
    \includegraphics[width=0.99\textwidth]{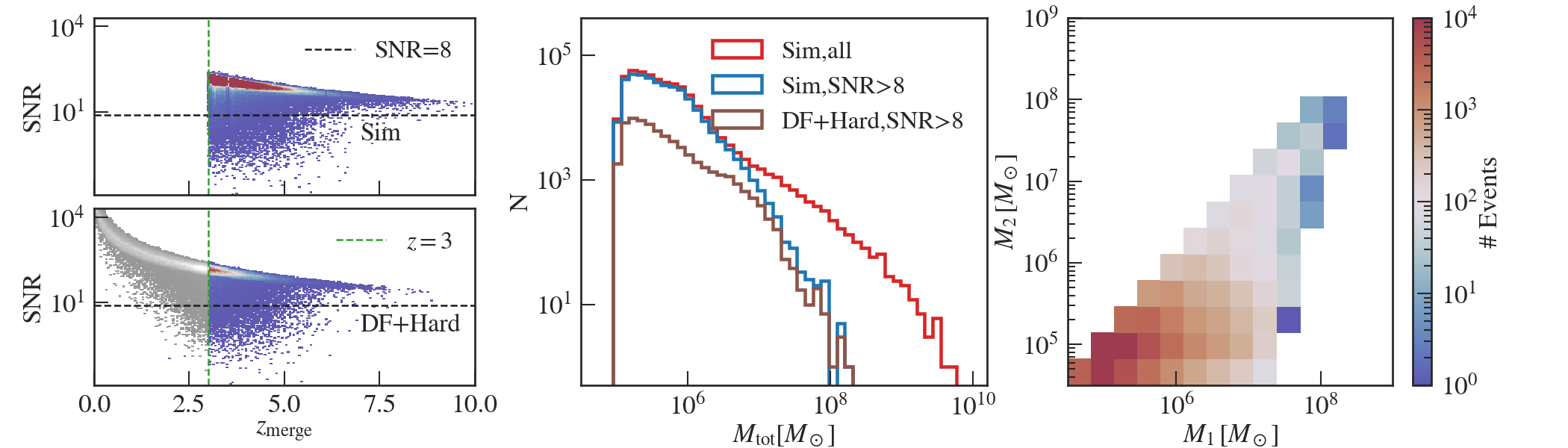}
    \caption{\textbf{Left:} the joint distribution of the SNR and redshift for \astrid~mergers. The top row is the SNR computed before the DF and hardening delays, and the bottom row is the SNR after the delay time is applied. 
    The mergers delayed to $z<3$ are masked in grey.
    \textbf{Middle:} distribution of binary mass for all \astrid~mergers (red), the ones with SNR>8 without the delay model (blue), and the ones that merge before $z=3$ after the delays (brown). The SNR>8 cut eliminates all mergers with $M_{\rm tot}>10^8 M_\odot$, while the drop in low-mass merger events is due to the delays. 
    \textbf{Right:} the distribution of two MBH masses for LISA detectable merger events at $z>3$. 
    Most events are expected to involve two seed-mass MBHs.}
    \label{fig:snr}
\end{figure*}

Although the strain in Figure \ref{fig:hc_fc} is a good estimation of the detectibility of a circular binary, for the eccentric case a more careful prediction comes from the signal-to-noise ratio (SNR). The SNR is estimated by integrating the ratio of the signal to the noise in the frequency domain. The sky, orientation, and polarization averaged SNR is given by :
\begin{equation}
\label{eq:snr}
\langle {\rm SNR} \rangle^2 = \frac{16}{5}\int_{f_{\rm start}}^{f_{\rm end}} \frac{h_s^2}{h_N^2} f^{-1} df,
\end{equation}
where $f_{\rm start} = f(t_{\rm start})$ and $f_{\rm end} = f(t_{\rm end})$, with $t_{\rm start}$ and $t_{\rm end}$ representing the starting and ending time of when the signal is observed. Note that here we are assuming eccentric waveforms for the binaries, and thus $h_s$ is given by the sum over different modes following Equation \ref{eq:ecc_hf}. As it is not computationally feasible to sum an infinite number of modes, we truncate the sum in Equation \ref{eq:ecc_hf} at $n=50$ and we have checked that the difference between the first $50$ and the first $100$ modes is less than $5\%$.

For the current configuration, we assume that the LISA observation lasts for 4 years. We further assume a most optimistic SNR for all mergers by taking $t_{\rm end}=t_{\rm peak}$ and $t_{\rm start}=t_{\rm peak}-4{\rm yrs}$. Under this assumption, we are always integrating the part of the waveform where the strain is maximized. However, as was discussed in \citet{Salcido2016} and \citet{Katz2020}, the actual SNR may be smaller if there is an offset between the LISA observation window and the merger time of the binary.

Figure \ref{fig:snr} shows the distribution of the SNR computed for all mergers in the simulation. 
The left column shows the joint distribution of SNR and the merging redshift. 
The top row is the SNR computed with merging redshifts before the DF and hardening delays, and the bottom row is the SNR after the delay time is applied. 
As was expected from the simpler calculation shown in Figure $\ref{fig:hc_fc}$, the majority of the binary population in the simulation has a SNR larger than the LISA detection threshold of 8 (plotted as dashed gray lines). 
The ones that falls below the SNR cut are mainly massive mergers with $M_{\rm tot}>10^7\,M_\odot$.
When we account for the delays, the mergers are pushed towards lower redshifts, and the resulting SNR is higher for each event.
However, when restricted to $z>3$, the 

The middle panel shows the effect of delays and the SNR cut on mergers with different masses.
The SNR cut removes all mergers with $M_{\rm tot}>10^8\,M_\odot$ from the LISA-detectable population.
On the low-mass end of MBH mergers, the reduction results from the DF and binary hardening delays.
Combining both the delays and SNR cut, we see that the overall detectable mergers at $z>3$ are $\sim 15\%$ of the original \astrid~merger population across all masses.
The seed-mass mergers still dominate over other events even though they are most strongly suppressed by the delays. 
Finally, on the right panel we show the mass distribution of the two MBHs involved in each detectable events.
The majority of these events are expected to be mergers from two seed-mass MBHs.
On the high-mass end, the detectable events has a mass ratio of $q\sim 1$ (close to the diagonal line).
Based on these results, the likelihood that a LISA detection comes from mergers of MBH seeds is high, but the detectable MBH seed mergers is only a small sample of the seed MBH pairs and the associated galaxy mergers.

Here, accounting for the delay time to merger affects the resulting SNR more than the eccentricity. 
The eccentricity itself, however, may affect the prospects for multi-messenger follow-up.
For example, eccentric binaries may spend a shorter amount of time in the LISA band compared to circular binaries. 
Spin-orbit interactions in eccentric binaries may change the orbital inclination with respect to the line-of-sight, which may also play a role in detectability and sky localization. 
We will explore such effects and their implications for multi-messenger follow-up in a companion paper.

\section{Conclusion and Discussion}
\label{sec:conclusion}
In this work we have made predictions for the MBH merger rate and associated LISA events for a cosmological population of MBHs with masses ranging between $5\times 10^4 M_\odot$ and $10^{10} M_\odot$ down to $z=3$, using the large volume cosmological simulation \astrid. 
At high redshifts, MBH mergers and the associated GW signal should provide strong contraints for models of seed black hole formation. 
In \astrid~, MBH seeds range from $5\times 10^4 M_\odot$ to $5\times 10^5 M_\odot$, covering down to masses that LISA will be most sensitive too. 
Moreover, in \astrid~we have included an on-the-fly subgrid dynamical friction prescription, which allows us to trace the MBH orbits down to the resolution limit. 

Using the MBH orbits directly from the simulation, we estimated the orbital eccentricity for MBH pairs that undergo merger in \astrid~simulation. 
In addition, we use the most recent post-processing models to account for additional delay in MBH mergers due to dynamical friction \citep[][]{Dosopoulou2017} and binary hardening \citep[][]{Vasiliev2015} at scales not resolved directly by \astrid. 
This is done by accounting for the orbital eccentricities constrained by the simulation which is important for the loss-cone scattering and gravitational-wave hardening phase. 
After considering the effect of these processes in delaying the MBH merger, we made detailed prediction of the expected number of mergers down to $z=3$, the redshift that the simulation has currently reached. 
Finally, we computed the detectability of these events by LISA. 

We find that most MBHs pairs in \astrid~have eccentric orbits distributed near $\epsilon=0.8$. 
We verify the eccentricity measurements by using both the shape and the dynamical information of the MBHs and find general agreement on the result. 
While some orbits circularize during the dynamical friction decay, the majority of them still maintain a high level of eccentricity at the time of the numerical merger. 
The orbital eccentricity is important in accelerating the binary hardening process. 
In particular, we show that the assumption of circular orbits for all binaries leads to estimates for the binary hardening time that can exceed $20$ Gyrs for most \astrid~binaries. 
Taking into account the measured orbital eccentricities, our estimated hardening times fall between $1\sim10$ Gyrs.

Even after considering the accelerated binary hardening rate due to eccentric orbits, for \astrid~mergers close to the seed mass, the binary hardening (including LC and GW hardening) time typically provides the longest
delay, and it remains more important than the dynamical friction component (including DF time modeled in \astrid~directly and the estimated sub-resolution component). 
For MBH binaries above the seed mass, the hardening time becomes comparable to the DF time and always remains $<1$ Gyrs. 
By comparing the DF directly modeled in \astrid~with the post-processed (sub-resolution) DF time, we find that they are comparable, accounting for  $100\sim 300$ Myr of binary evolution. 
At the resolution of \astrid, the sub-grid DF added directly in the simulation is able to recover more than half of the dynamical friction decay process before the numerical merger. 

Without accounting for any additional post-processed binary dynamics delays, we expect $\sim 2$ merger events per year \citep[][]{DeGraf:2021inprep} from the $z>3$ MBH population in \astrid.
With the post-processed dynamical friction and binary hardening taken into account, the expected merger rate reduces to $0.3\sim 0.7$ per year at $z>3$. 
\astrid~predicts for merger rates that are higher than most previous predictions from hydro-dynamical simulations of comparable size \citep[e.g.][]{Salcido2016,Katz2020,Volonteri2020}, because \astrid~accounts for a seed population \citep[see][for a more direct comparison]{DeGraf:2021inprep} in halos about an order of magnitude lower in mass than e.g. Illustris ($M_{\rm halo,thr} = 7 \times 10^{10}\, M_\odot$) and EAGLE ($M_{\rm halo,thr} = 1.4 \times 10^{10}\, M_\odot$).
Among the whole MBH merger population, the seed-mass mergers are most affected by the delays, with only $<20\%$ of the original simulation mergers still merging at $z>3$.
Nonetheless, because the seed-mass mergers dominate the merger population in absolute numbers (250455 out of 440999), they still occupy a large fraction of the delayed mergers.
Out of the delayed merger events at $z>3$, $\sim 60\%$ involve two seed-mass MBHs, $\sim 30\%$ are mergers between one non-seed MBH and one seed-mass MBH, and $\sim 10\%$ are mergers between two large mass MBHs.

We use a 4-year LISA observation time to calculate an upper limit on the SNR for each merger event. 
Many of these high-z mergers result in SNRs around $\sim 200$.
With a SNR>8 threshold, high-mass merger ($M_{\rm tot}>10^8\,M_\odot$) events are removed from the detectable population at $z>3$.
The $M_{\rm tot}<10^7\,M_\odot$ mergers are still detectable.
As a result, the LISA detectable population is still dominated by seed MBH mergers, and the expected detection rate is similar to the total merger rate of $0.3\sim 0.7$ per year at $z>3$.

Based on these results, a LISA detection of merger events from MBH seeds population is highly feasible.
However, the detectable MBH seed mergers are predicted to correspond to the sample of the seed MBH pairs that occur in hosts with stellar masses close to $10^{9} \msun$. 
This is about three times larger than the typical stellar mass at which seed-mass mergers are expected to occur if loss cone scattering was not accounted for.
We also find that $\sim 80\%$ of the seed-seed merger remnants in the simulation are the only MBH residing in their host galaxies.
Accounting for the DF and binary hardening delays slightly favors systems embedded in a larger galaxy with a more massive MBH around.
This is because the more massive hosts tend to provide a higher stellar density and hence a more effective loss-cone scattering.
However, sole MBH remnants still make up $\sim 70\%$ of the seed-seed merger population after the delays.
Regardless, \astrid~predicts the host galaxies of the detectable $z>3$ mergers to be galaxies of $M_* \sim 10^9-10^{10}\msun $. 
These host galaxies are detectable with current and upcoming telescopes.


We note also that our estimation of the low-mass MBH merger rate is a lower-limit, since we do not resolve the MBHs residing in low-mass dwarf galaxies.
Observations have provided evidence that dwarf galaxies host MBHs in their center \citep[e.g.][]{Reines2013,Moran2014,Satyapal2014,Lemons2015,Sartori2015,Pardo2016,Nguyen2019}.
Simulations \citep[e.g.][]{Wassenhove2010,Bellovary2019,Volonteri2020} also shows that dwarf galaxies consistently merge into larger galaxies over time. 
Hence, missing the dwarf galaxy MBHs could bias our merger rate and detection rate estimation towards the lower end.

We find that current simulations such as \astrid~are getting closer to predicting DF timescales for the binary evolution, but the estimation of the binary hardening timescale remains more uncertain as it depends on the properties of central stellar densities below the resolution limit.
We have shown that changing the stellar density extrapolation starting point from $1.5\epsilon_g$ to $2\epsilon_g$ increases the estimated density at the influence radius by a factor of $\sim 10$, and thereby shortens the estimated binary hardening timescale by a factor of $\sim 10$. 
This translates to a factor of $\sim 3$ different in the merger rate predictions. 
To more confidently estimate the binary hardening timescale and thus the MBH merger rate in the context of cosmological simulations, a better modeling of the inner region of the galaxy would be needed. 
Nonetheless, we still expect the merger rates to be within a factor of a few of what a cosmological simulation is able to predict (at the resolution of \astrid)

Finally, in this work, we do not evolve the orbital eccentricity during the loss-cone scattering phase. 
Loss-cone scattering can increase the orbital eccentricity of the binary \citep[e.g.][]{Sesana2010,Kelley2017b}, and may affect the detected GW signal.
We also do not consider circumbinary-disk interactions \citep[e.g.][]{Haiman2009}, since circumbinary-disk simulations for eccentric binaries have not yet been comprehensively explored for a wide-enough range of binary parameters and disk properties.
A significant amount of progress, however, has been made in the hydrodynamic modeling of such systems \citep[e.g.,][]{Duffell2020,Tiede2020,DOrazio2021}.
Binary-disk interactions may also affect the spin orientations of each MBH. 
It is also currently uncertain how a circumbinary disk would respond when an eccentric binary undergoes post-Newtonian spin-orbit interactions.
We thus leave such analyses with our cosmological binary population for future work. 

\section*{Acknowledgements}
\astrid~was run on the Frontera facility at the Texas Advanced Computing Center.

TDM and RACC acknowledge funding from 
the NSF AI Institute: Physics of the Future, NSF PHY-2020295, 
NASA ATP NNX17AK56G, and NASA ATP 80NSSC18K101. TDM acknowledges additional support from  NSF ACI-1614853, NSF AST-1616168, NASA ATP 19-ATP19-0084, and NASA ATP 80NSSC20K0519, and RACC from NSF AST-1909193.
SB was supported by NSF grant AST-1817256. 
AMH is supported by the McWilliams Postdoctoral Fellowship.

\section*{Data Availability}

The code to reproduce the simulation is available at \url{https://github.com/MP-Gadget/MP-Gadget}, and continues to be developed. Text file forms of the data presented here as well as scripts to generate the figures are available. Binary catalogs including the MBH information and the host galaxy properties are available upon request.

\bibliographystyle{mnras}
\bibliography{bib.bib}

\end{document}